\documentclass{article}
\usepackage{graphicx} % Required for inserting images
\usepackage{float} % Required for inserting images
\usepackage{amsmath} % Required for math
\usepackage{booktabs} % Required for table
\usepackage{subcaption} % Required for subfig
\usepackage{amssymb} % Required for symbol
\usepackage{hyperref} % Required for hyper link
\usepackage[utf8]{inputenc}

\title{Building a Life Cycle Assessment Model using Bayesian Networks}
\author{Cedric Fraces Gasmi, Wennan Long}
\date{November 2023}

\begin{document}

\maketitle
\begin{abstract}
This paper introduces the Oilfield Pollutant Graphical Model (OPGM), an innovative approach designed to improve the benchmarking and uncertainty analysis of greenhouse gas (GHG) emissions in oilfields. Building on the robust foundation provided by the Oil Production Greenhouse Gas Emission Estimator (OPGEE) framework, OPGM retains all essential functionalities of the latest OPGEE iteration (v3.0c), while offering substantial improvements in user experience and computational performance. Key advances of OPGM include a streamlined user interface for more intuitive interaction, which facilitates transparent visualization of intermediate results and thus contributes to a more interpretable and accessible analysis process. A notable feature of the OPGM is its ability to naturally perform sensitivity analyzes. This is achieved by allowing users to seamlessly transition nodes from deterministic to probabilistic, thereby integrating uncertainty directly into the core structure of the model. OPGM achieves remarkable computational efficiency, executing analyzes at a speed 1e+5 times faster than the Excel-based OPGEE, thus facilitating rapid large-scale emissions assessments. This leap in processing speed represents a significant step forward in emissions modeling, enabling more agile and accurate environmental impact assessments. The integration of OPGM into existing Life Cycle Assessment (LCA) practices holds the promise of significantly improving the precision and speed of environmental impact analyses, offering a vital tool for policymakers and industry stakeholders in their efforts to better understand and manage the environmental impacts of oilfield operations.
\end{abstract}

\section{Introduction}
Life Cycle Assessment (LCA) is a systematic approach to evaluating the environmental impacts of a product or process throughout its entire life cycle, from the extraction of raw materials to the disposal of waste. Oil fields are complex systems with a wide range of environmental impacts, including greenhouse gas emissions, water pollution, and land degradation. According to the International Energy Agency (IEA)~\cite{iea2023emissions}, oil and gas production, transport and processing resulted in 5.1 billion tonnes (Gt) of equivalent CO2 (CO2eq.) in 2022 - just under 15\% of global energy sector GHG emissions. LCA is a valuable tool for understanding and managing these impacts. We propose a bottom-up engineering-based emission modeling to evaluate the scope 1, scope 2, and scope 3 emissions of an oil operation from exploration to extraction development operations, surface handling and transport.

Probabilistic Graphical Models (PGM) (also called Bayesian Belief Networks~\cite{Pearl1988}) can be used to represent complex systems and make predictions about their behavior~\cite{koller2009probabilistic}. We present a new LCA model (OPGM) built using the PGM library TKRISK\textsuperscript{\textregistered}. We show that the PGM can accurately capture the complexity of a well-established reference model while opening each of its components to the layman's eyes. We compare the metrics computed using our PGM with those of the original model on a benchmark of over 1,000 fields. We show that OPGM presents a scalable alternative to existing solutions and can be used to perform advanced evaluations, including sensitivity analysis and uncertainty quantification.

\section{Related work}
The Oil Production Greenhouse Gas Emission Estimator (OPGEE) is an engineering-based life cycle assessment (LCA) tool that is used primarily to estimate upstream greenhouse gas (GHG) emissions. OPGEE is recognized for its bottom-up LCA approach and has gained widespread adoption across various sectors. Regulators, operators, consulting firms, financial institutions, autonomous intergovernmental organizations, and non-governmental organizations (NGOs) have implemented OPGEE in their operations~\cite{OPGEE_v3}. Among its users are the California Air Resources Board (CARB) under the Low Carbon Fuel Standard (LCFS), Chevron, McKinsey \& Company, S\&P Global Platts, the International Energy Agency (IEA) and the Rocky Mountain Institute (RMI).

OPGEE is designed to simulate upstream GHG emissions across four LCA stages. Exploration \& Development, Production, Surface Processing, and Transportation. It can model various production methods such as waterflooding, gasflooding, steamflooding and gas lifting, along with common surface operations such as venting, flaring, and water \& gas reinjection or disposal \cite{OPGEE_v3}\cite{Hassan2013}\cite{MASNADI_proxy} \cite{masnadi2018}.

The latest iteration, \emph{OPGEE v3.0c}, includes significant enhancements, particularly in the fugitive emission model. These improvements ensure a better representation of the GHG emissions from all oil and gas operations. A notable advancement in this version is the incorporation of a component-level fugitive model supported by a comprehensive database of component-level activity and emissions measurements \cite{Rutherford2021}.

To our knowledge, this is the first time an LCA model of such an extent was built entirely as a PGM.

\section{Methodology Overview}
\begin{figure}[H]
    \centering
    \includegraphics[width=1\linewidth]{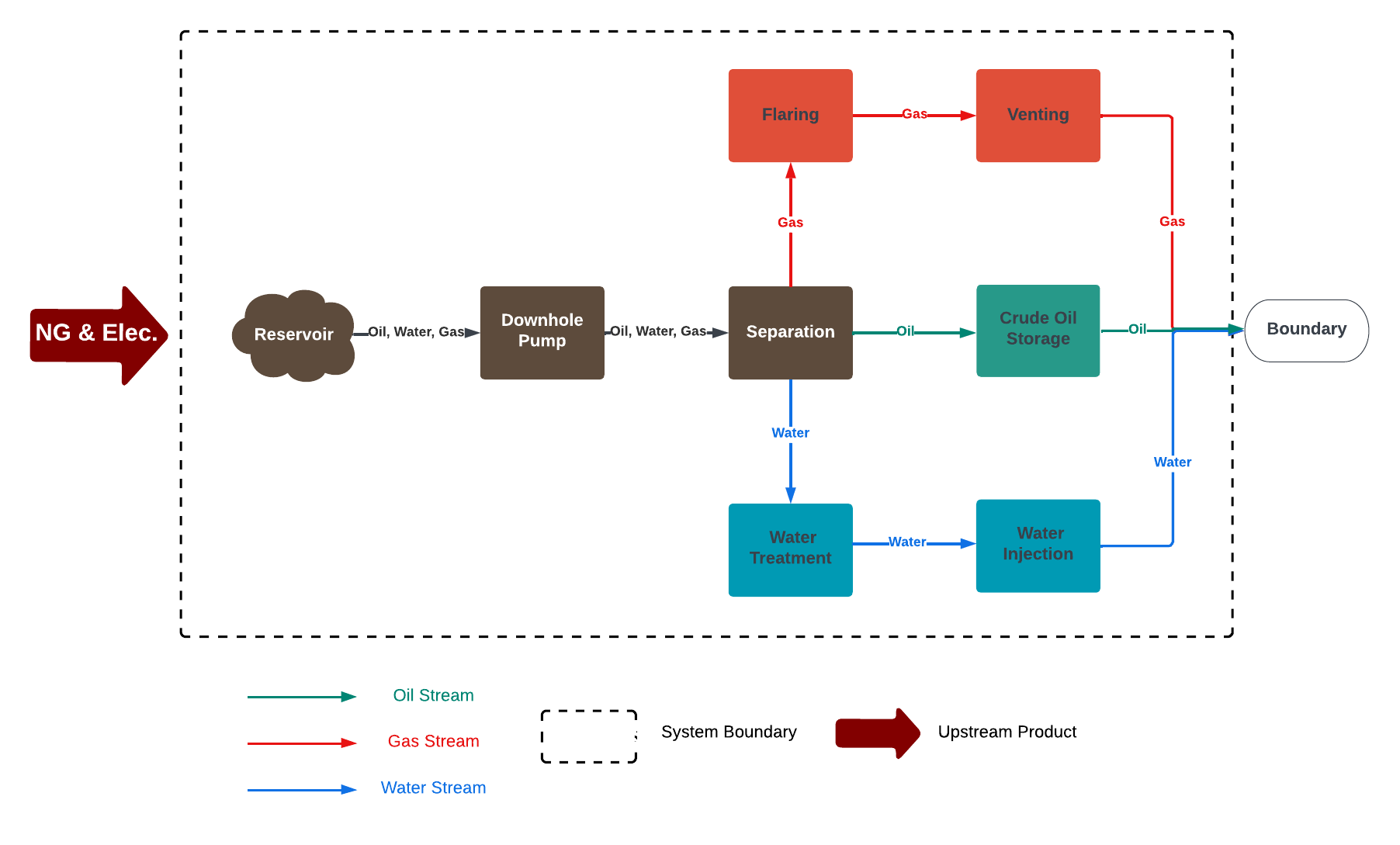}
    \caption{Definition of oil and gas system's boundary for LCA methodology}
    \label{fig:flowchart}
\end{figure}

We propose to model the carbon intensity (CI) of an oilfield's operation from extraction to refinery. The boundaries of the system are represented in Figure~\ref{fig:flowchart}

The model operates under the assumption that three fluids are present in the system:
\begin{enumerate}
    \item Oil Stream: Oil is produced from the reservoir, passes through a downhole pump, and then is transferred to a separation process where oil, gas, and water are separated. Separated oil is stored in the crude oil storage tank and transported outside the system's boundary for further processing or shipping.

    \item Gas Stream: Gas from the separation process is directed to flaring, where excess gas is burned off, and to venting, where gas is released directly into the atmosphere without being burned.

    \item Water Stream: Water from the separation process is sent to a water treatment facility. After treatment, the water is either disposed of or used in water injection processes, likely for enhanced oil recovery or to maintain reservoir pressure.
\end{enumerate}

We represent the LCA model as a PGM. PGMs(\cite{koller2009probabilistic}) are a class of statistical models that represent conditional dependencies between variables using a graphical structure that allows reasoning and inference under uncertainty.

\begin{figure}[H]
    \centering
    \includegraphics[width=1\linewidth]{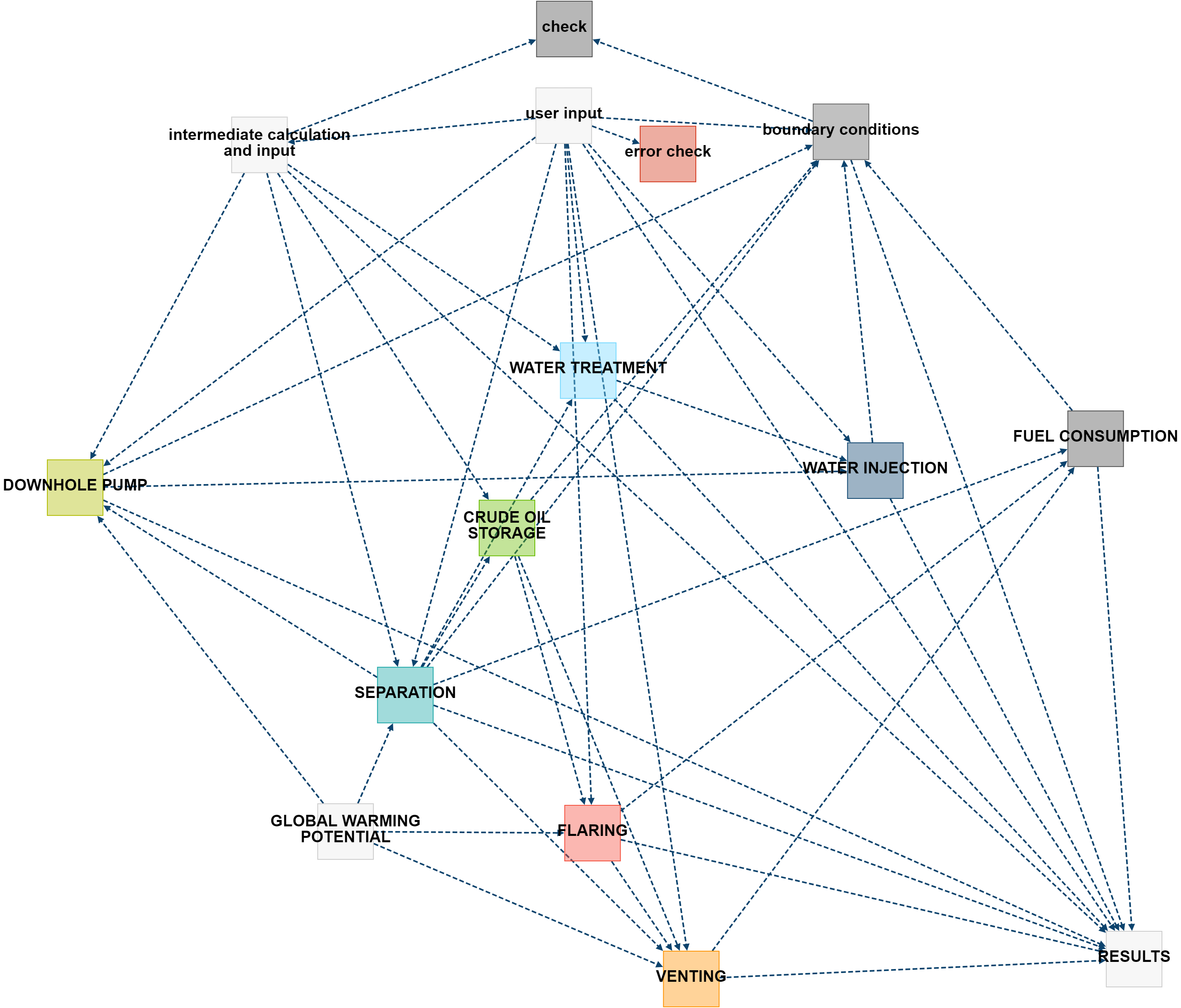}
    \caption{Graphical representation of LCA-OPGM main modules and their relationships (\emph{TKRISK}$^{\tiny{\textregistered}}$).}
    \label{fig:OPGEE_FULL}
\end{figure}

Key properties of PGMs include:
\begin{enumerate}
    \item Graphical representation: PGMs employ graphs to visually depict the relationships between random variables. A node in the graph represents each variable, and the edges connect nodes with direct dependencies.
    \item Probabilistic framework: PGMs utilize probability theory to quantify the strengths of these dependencies. Probability distributions are associated with each variable and edge, representing the likelihood of different values given the values of its parents or children.
    \item Directed acyclic graphs (DAGs): restricts the graph structure to ensure that no directed cycles exist. This constraint ensures that there is no circular dependency among variables, making the model well defined and computable.
\end{enumerate}
A PGM is a knowledge integration tool deployed to manage uncertainty and support decision-making processes. It can also be considered as an intuitive graphical representation of dependencies relationship between different indicators and variables. These dependencies are quantified using probability, hence making possible to use PGMs to compute the probability of an event given evidence on the other variables studied, which could be both hard (data) or soft (expert knowledge) evidences. 
TKRISK\textsuperscript{\textregistered} (www.teokononda.com) is a Bayesian network framework that can be used to build risk models. The graph-based LCA model (OPGM) is thought of and built as a risk model. Figure \ref{fig:OPGEE_FULL} shows a graphical representation of the model.

The nodes presented in Figure~\ref{fig:OPGEE_FULL} are actually collapsed groups. We present each of these groups over the next sections.

\subsection{User input}
In this section, we present the various parameters collected through the user interface.

\begin{figure}[H]
\centering
\includegraphics[width=1\linewidth]{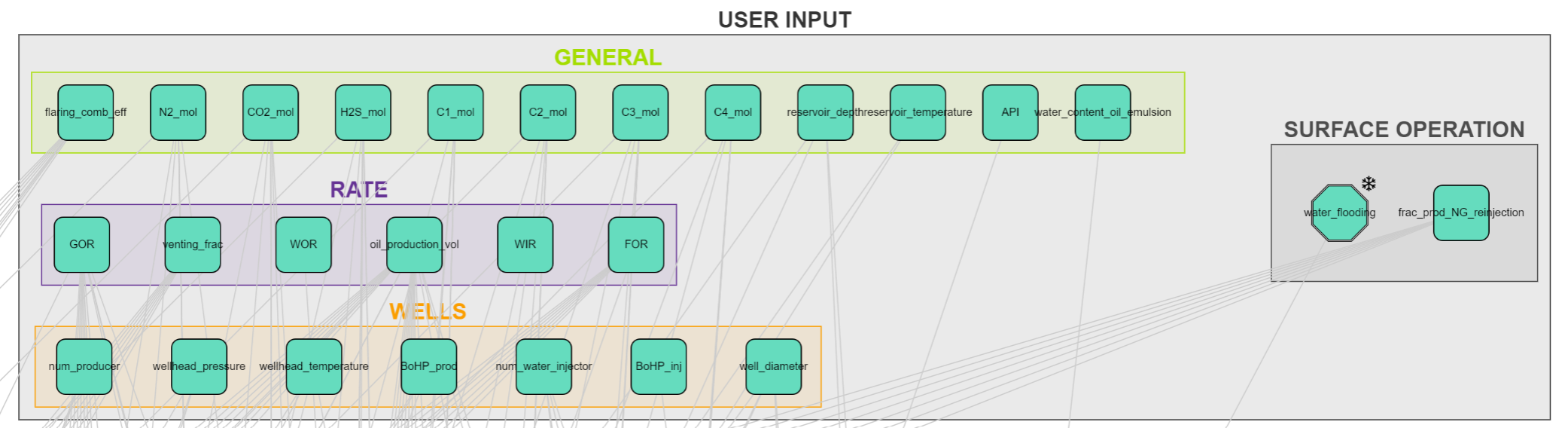}
\caption{User input}
\label{fig:user_input}
\end{figure}

Figure \ref{fig:user_input} presents the user input panel. Input includes -but is not limited to variables related to surface operation methods, production rates, and specific well information. Fluid parameters such as the gas molecular fraction and API gravity are also incorporated. The current version contains modules for the disposal and reinjection of water.

\subsection{Model output}
\begin{figure}[H]
    \centering
    \includegraphics[width=0.5\linewidth]{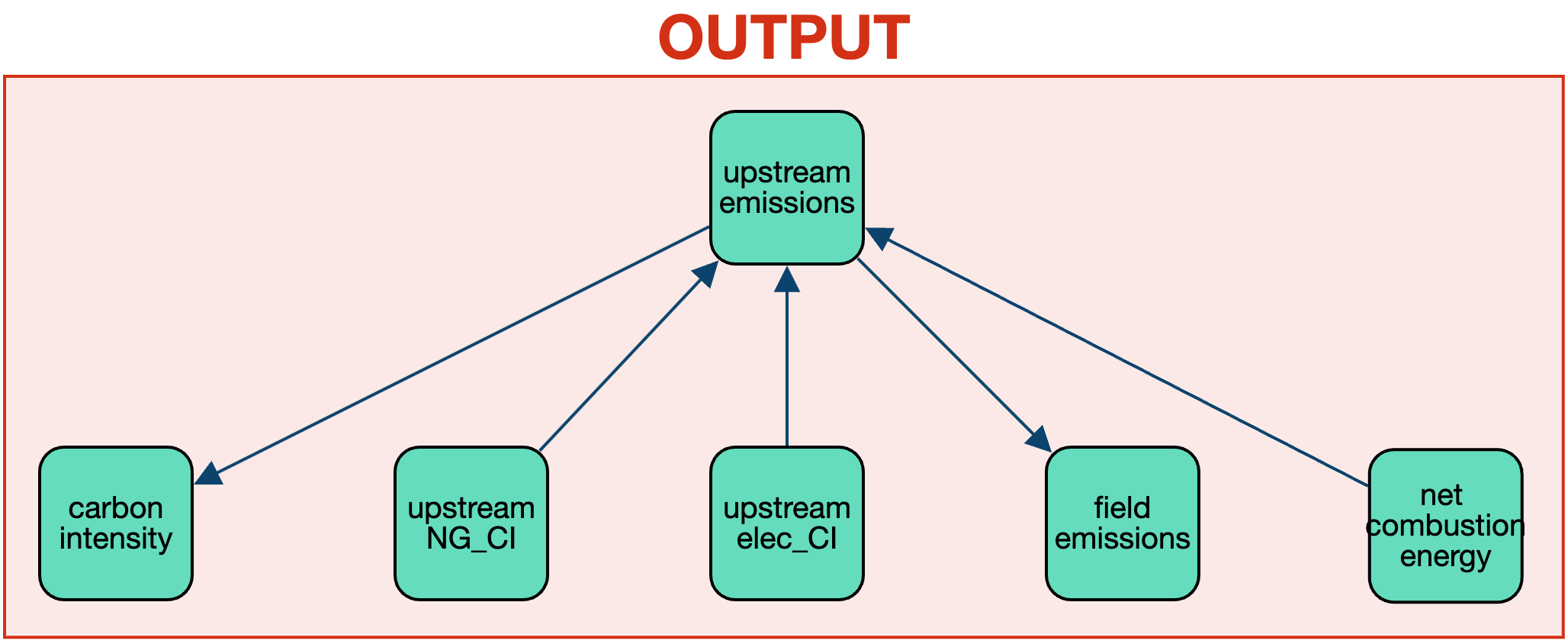}
    \caption{Output panel of OPGM}
    \label{fig:ghg_opgee_results}
\end{figure}

The model outputs a series of KPIs related to GHG emissions. Figure~\ref{fig:ghg_opgee_results} shows these output metrics.

\begin{itemize}
    \item Carbon Intensity: kgCO2eq./BOE
    \item Upstream NG CI: Upstream Natural Gas GHG carbon intensity (gCO2eq./mmbtu)
    \item Upstream elec CI: Upstream Electricity GHG carbon intensity (gCO2eq./mmbtu)
    \item Upstream Emission: imported upstream emissions (tCO2eq./day)
    \item Field Emissions: field total emissions (tCO2eq./day)
    \item Net Combustion Energy: The total combustion energy demand minus LHV of the gas produced (mmbtu/day)
\end{itemize}

\subsection{Downhole pump}
Artificial lift systems are ubiquitous in oil and gas production. They facilitate the movement of fluids from underground reservoirs to the surface. These systems contribute significantly to greenhouse gas (GHG) emissions, primarily through fugitive emissions and the energy consumed in pumping. Our model includes a downhole pump module designed to calculate the energy use and associated emissions from artificial lift operations. This module comprises five key components: wellbore conditions, bottom-hole fluid properties, well-bore fluid properties, mass rate, and energy usage and emissions. In particular, energy usage and emissions, along with the mass rate component, are integral to all modules in the system.

\begin{figure}[H]
\centering
\includegraphics[width=1\linewidth]{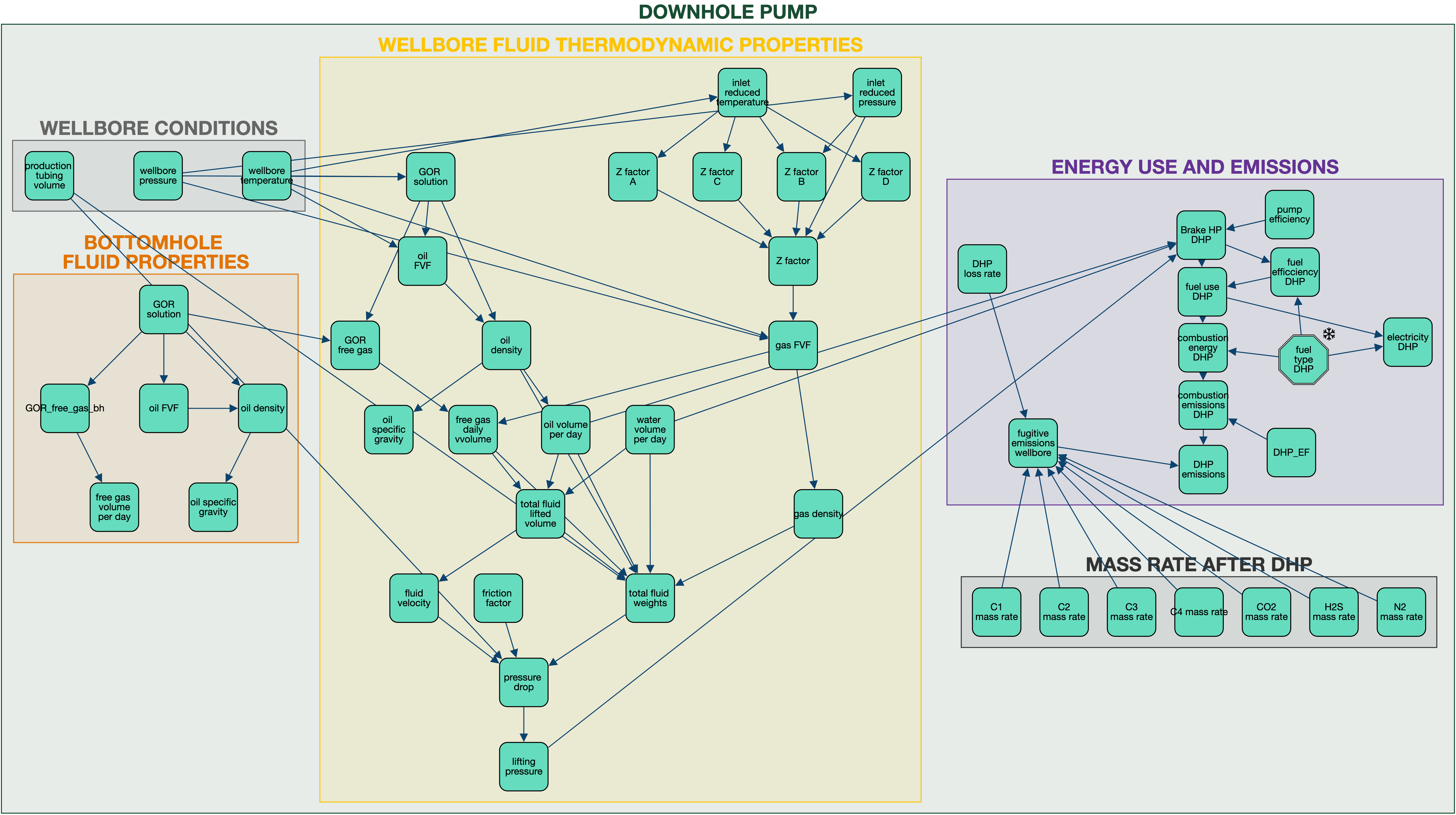}
\caption{Schematic of the Downhole Pump Module}
\label{fig:downhole_pump}
\end{figure}

The module includes the following calculations:

\begin{itemize}

    \item Pressure and Energy Calculations: The energy required to lift fluids is calculated based on overcoming the pressure traverse, which includes resistance to gravity flow and frictional losses. The pressure required is the sum of the wellhead pressure, the pressure traverse, and subtracting wellbore pressure.

    \item Fluids and Properties: This part involves calculating the average properties of the fluids within the wellbore, such as pressure, temperature, and density. These calculations take into account variables like the gas-oil ratio, as well as wellhead and wellbore conditions.

    \item Well Pressure Traverse: This involves calculating the gravitational head, which is the pressure drop due to gravity, and the frictional pressure drop resulting from fluid contact with the well tubing. The calculation uses fluid density, wellbore volume, and tubing cross-sectional area.

    \item Pump Power and Efficiency: The brake horsepower of a pump is calculated using its discharge flow rate, pumping pressure, and efficiency. The pumping pressure differs between the pump discharge and the suction pressures.
\end{itemize}

\subsection{Separation}
\label{sec:Separation}
This stage involves the separation of the fluids produced into their respective components, primarily oil, gas, and water. This process takes place in a pressure vessel called a separator, which is designed to utilize various physical principles to achieve effective separation. One of the primary principles employed in the separation process is gravity. As the fluids produced enter the separator, they are allowed to settle under the influence of gravity. This allows heavier components, such as water and solids, to sink to the bottom of the separator while lighter components, such as oil and gas, rise to the top. This initial separation is aided by using internal baffles and coalescing plates, which help to entrap and enlarge small droplets, allowing them to settle more efficiently.

Another important principle involved in the separation process is the reduction of pressure. As the fluids produced flow into the separator, the pressure is gradually reduced. This pressure drop causes the release of dissolved gases, primarily methane, from the liquid phase. These gases rise to the top of the separator and are collected separately.

\begin{figure}[H]
    \centering
    \includegraphics[width=1\linewidth]{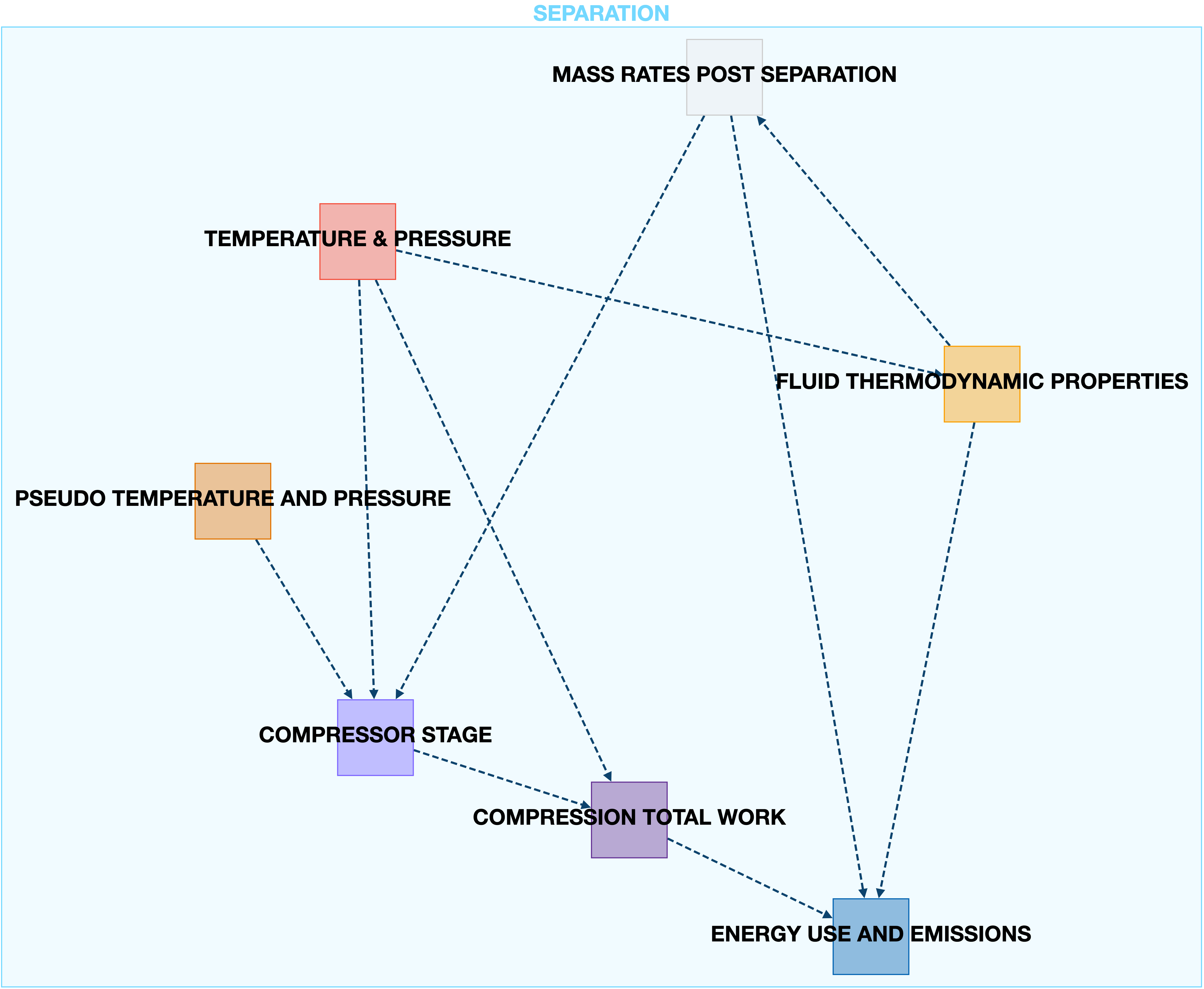}
    \caption{Separation module graph}
    \label{fig:separation_col}
\end{figure}

Figure~\ref{fig:separation_col} shows the main components of the separation module.

\begin{figure}[H]
    \centering
    \includegraphics[width=1\linewidth]{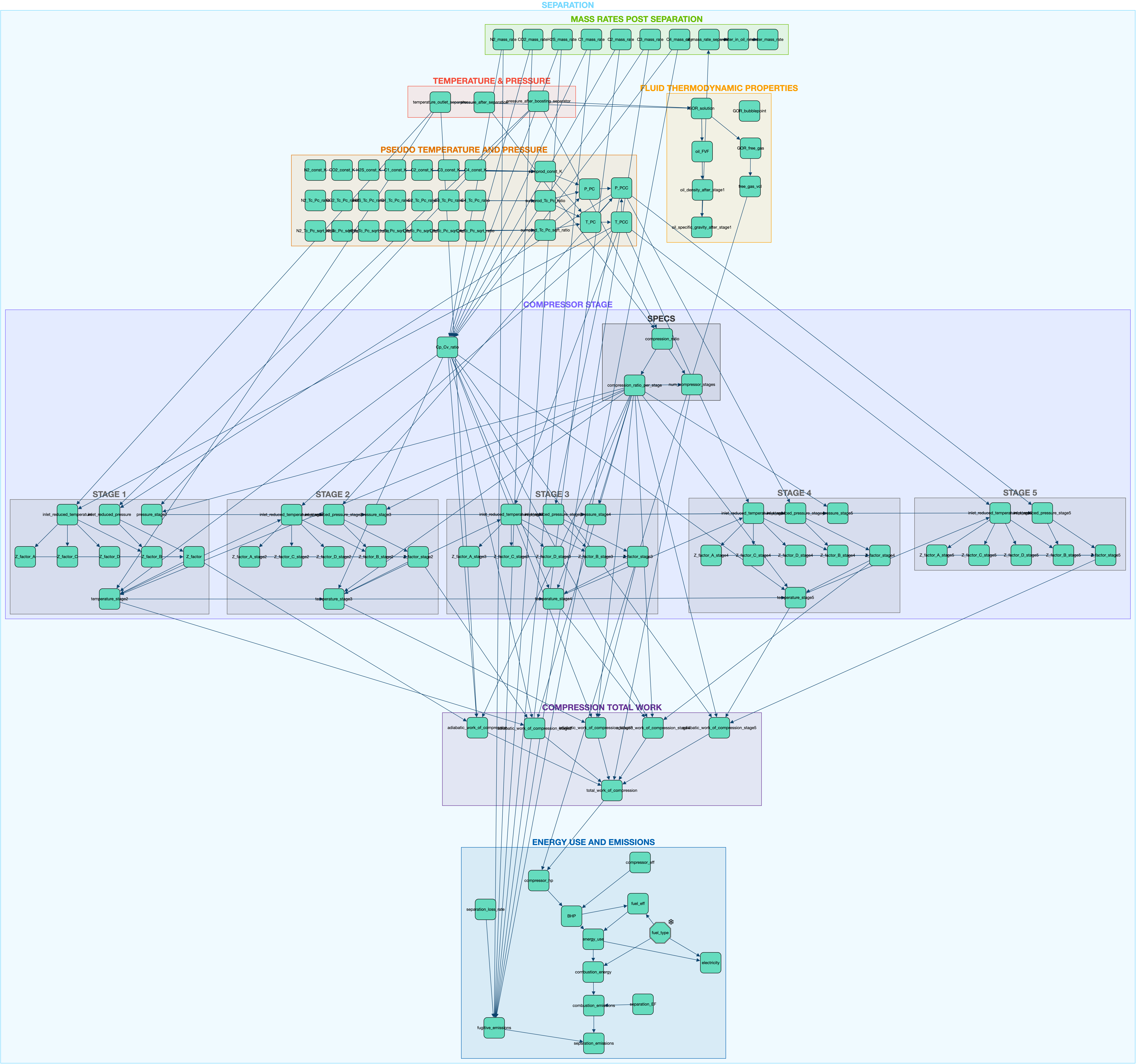}
    \caption{Separation module graph (expanded view)}
    \label{fig:separation}
\end{figure}

 An expanded view is represented in figure~\ref{fig:separation}

\begin{itemize}
    \item Separator Specifications: Specifications are based on package-built separators, with working pressures ranging from 720 to 1,440 psig. The capacity for liquid and gas processing varies with pressure. The selection of separators is based on the closest effective wellhead pressure.

    \item Gas-Specific Gravity and Heating Value: These parameters are assumed to be constant for all phases of separation. Gas-specific gravity is defined as a working parameter for use in later calculations.

    \item Separation Stages and Pressure: The solution gas oil ratio (Rs) at each separation stage is determined using a defined function, which accounts for the conditions at each stage.

    \item Free Gas Removal Calculation: The free gas removed at each separation stage is computed. This involves calculating the difference in gas in solution before and after each stage.

    \item Oil Properties at Each Stage: The oil formation volume factor (FVF) and the specific gravity are calculated at each stage, considering the varying amount of gas in solution.

    \item Compression Ratio and Energy Consumption: The compression ratio for each stage is calculated as the ratio of the outlet gas pressure to the stage operating pressure. On this basis, the work of the compressor and the energy consumption are calculated, similar to the calculations for individual compressors.

\end{itemize}

\subsection{Flaring}

Flaring is a method used to dispose of associated natural gas produced during bulk separation (see Section \ref{sec:Separation}) when its economic use is not feasible. Flaring contributes significantly to the GHG emissions from global oil fields \cite{masnadi2018}. In 2008, gas flaring resulted in the emission of approximately 0.28 GtCO$_2$ eq, accounting for approximately 1\% of global GHG emissions \cite{Elvidge2009}. Since 1994, the NOAA National Geophysical Data Center has been estimating flaring volumes using satellite imagery \cite{Elvidge2009}. The estimated flaring volumes by country show a highly skewed distribution \cite{Elvidge2009}.

\begin{figure}[H]
    \centering
    \includegraphics[width=1\linewidth]{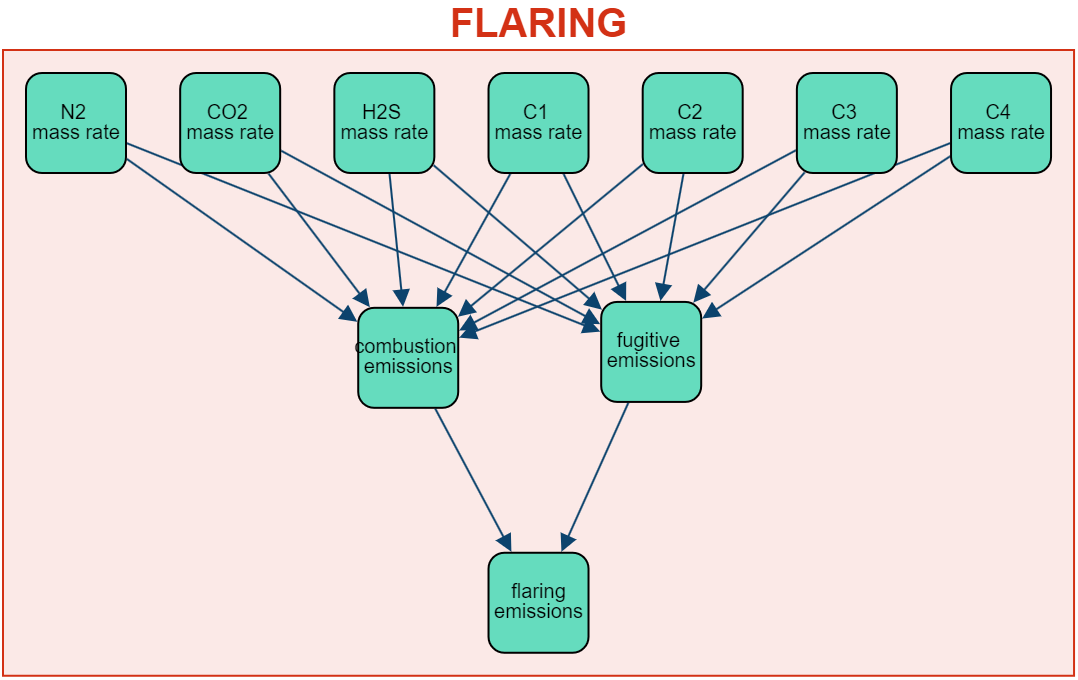}
    \caption{Flaring module}
    \label{fig:ghg_flaring}
\end{figure}
\begin{itemize}
    \item Flaring-to-Oil Ratio (FOR): The FOR, expressed in scf/bbl, is a crucial input parameter. It is converted to flaring volume using the volume of oil produced. The formula to calculate the flaring volume (QF) considers the FOR, the oil production volume (Qo), and additional flaring sources from oil storage tanks and the collection of off-gas from the mines.

    \item Flare Efficiency: Flare efficiency, which affects the amount of gas combusted versus emitted as unburned hydrocarbons, varies on the basis of several factors, such as flare exit velocities, wind speed, and gas composition. The OPGEE model uses empirical measurements from various studies to determine the efficiency of methane flare destruction.

    \item Combustion and Uncombusted Emissions: Emissions are divided into two categories: emissions from uncombusted gas and emissions from gas undergoing combustion. The former assumes that the gas slips through without reacting, while the latter assumes complete oxidation to CO2. 

    \item Total Flaring Emissions: Total flaring emissions are calculated as the sum of emissions from the stripped (uncombusted) and combustion processes.

\end{itemize}

\subsection{Venting}
Venting is the controlled release of unburned gases into the atmosphere. These gases can include natural gas, methane, and other hydrocarbon vapors, as well as water vapor and other non-hydrocarbon gases. Venting occurs during various stages of oil and gas production.
\begin{figure}[H]
    \centering
    \includegraphics[width=1\linewidth]{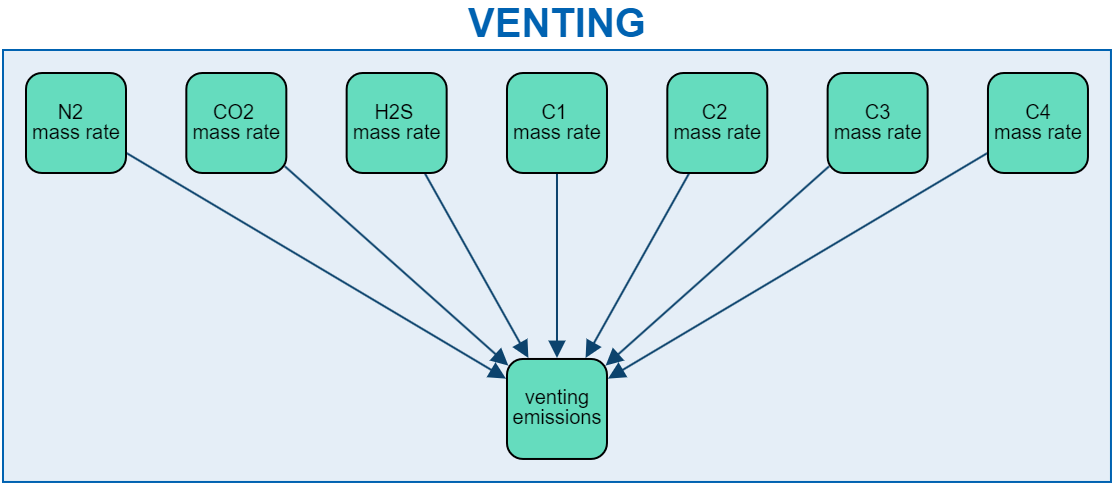}
    \caption{Venting module}
    \label{fig:venting}
\end{figure}
Venting is modeled in OPGEE based on the following considerations:

\begin{itemize}
    \item Equipment Maintenance and Safety: Venting is often used for safety reasons, such as depressurizing equipment for maintenance or to relieve unexpected pressure build-ups that could pose safety risks.
    \item Operational Necessity: In certain operational scenarios, such as well testing or equipment failures, venting may be used to manage gas that cannot be processed or sold.
    \item Lack of infrastructure: In some cases, especially in remote or older fields, the lack of infrastructure for gas capture or utilization leads to venting as a means of disposing of excess gas.

\end{itemize}

\subsection{Water treatment}
Oil production often generates substantial amounts of water, which can be contaminated with hydrocarbons, salts, and other unwanted constituents. A typical profile of these pollutants, which varies according to factors such as reservoir geology, is presented in Table \ref{tab:water_quality} \cite[p. 59]{Vlasopoulos2006}. The volume of water produced is determined by the Water-Oil Ratio (WOR). After treatment, this produced water is either reinjected, discharged into the local environment, or injected into aquifers.

\begin{table}[h]
\centering
\begin{scriptsize}
\caption{Typical Concentrations of Process Water Pollutants \cite{Vlasopoulos2006}}
\label{tab:water_quality}
\begin{tabular}{@{}cc@{}}
\toprule
Pollutants & Concentration (mg/l) \\ \midrule
Oil and grease & 200 \\
Boron & 5 \\
Total dissolved solids (TDS) & 5000 \\
Sodium & 2100 \\ \bottomrule
\end{tabular}
\end{scriptsize}
\end{table}

Process water from oil production can undergo a variety of treatments. OPGEE categorizes these technologies into four stages, as detailed in Appendix \ref{sec:water_treatment_table} Table \ref{tab:stages} \cite{Dillon2003}. This categorization and the energy consumption data for each technology, measured in kWh per cubic meter of water input, have been adopted from Vlasopoulos et al. \cite{Vlasopoulos2006}.

\begin{figure}[H]
    \centering
    \includegraphics[width=0.75\linewidth]{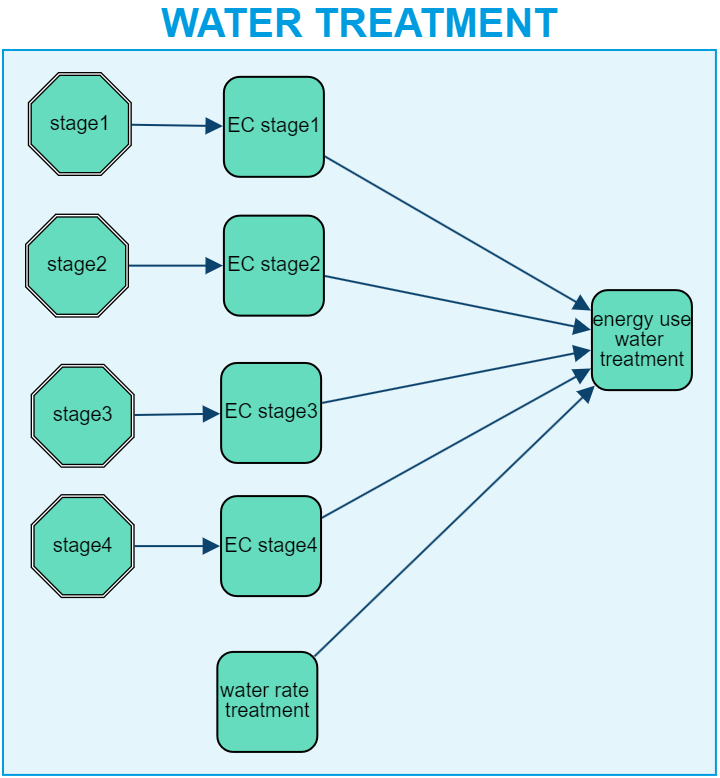}
    \caption{Water Treatment Module}
    \label{fig:water_treatment}
\end{figure}

Figure \ref{fig:water_treatment} illustrates the structure of the water treatment module in OPGEE. The operational workflow is described as follows:

\begin{itemize}
    \item \textbf{Treatment Train Setup:} Users can configure a treatment train in OPGEE by selecting a technology from each treatment stage.

    \item \textbf{Calculation of Electrical Consumption:} The total electricity consumption (\(E_{\text{tot}}\)) for water treatment is calculated based on the electricity consumption of each stage (\(e_{s,N}\)) and the water input into each stage (\(Q_{w, N}\)). OPGEE computes \(E_{\text{tot}}\) as the sum of the products of the electricity consumption per stage and the volume of water entering that stage.

    \item \textbf{Water Feed Calculation:} For the treatment of the produced water, the water feed for Stage 1 equals the flow of the well stream. Default volume losses are assumed to be 0\% for all technologies, except for wetlands, where a loss of 26\% is assumed. The water feed for the subsequent stages is adjusted to account for volume losses in the preceding stage.
\end{itemize}

\subsection{Water Injection}
Water disposal manages large volumes of wastewater generated during oil and gas production. This wastewater, also known as produced water, can contain various contaminants, including salts, hydrocarbons, and heavy metals. The most widely used method for disposing of the water produced is water injection. This method involves injecting the water produced into a deep underground formation that no longer produces oil or gas. Although this method is generally considered safe and environmentally friendly, it mobilizes large installations and requires external energy.

\begin{figure}[H]
    \centering
    \includegraphics[width=0.75\linewidth]{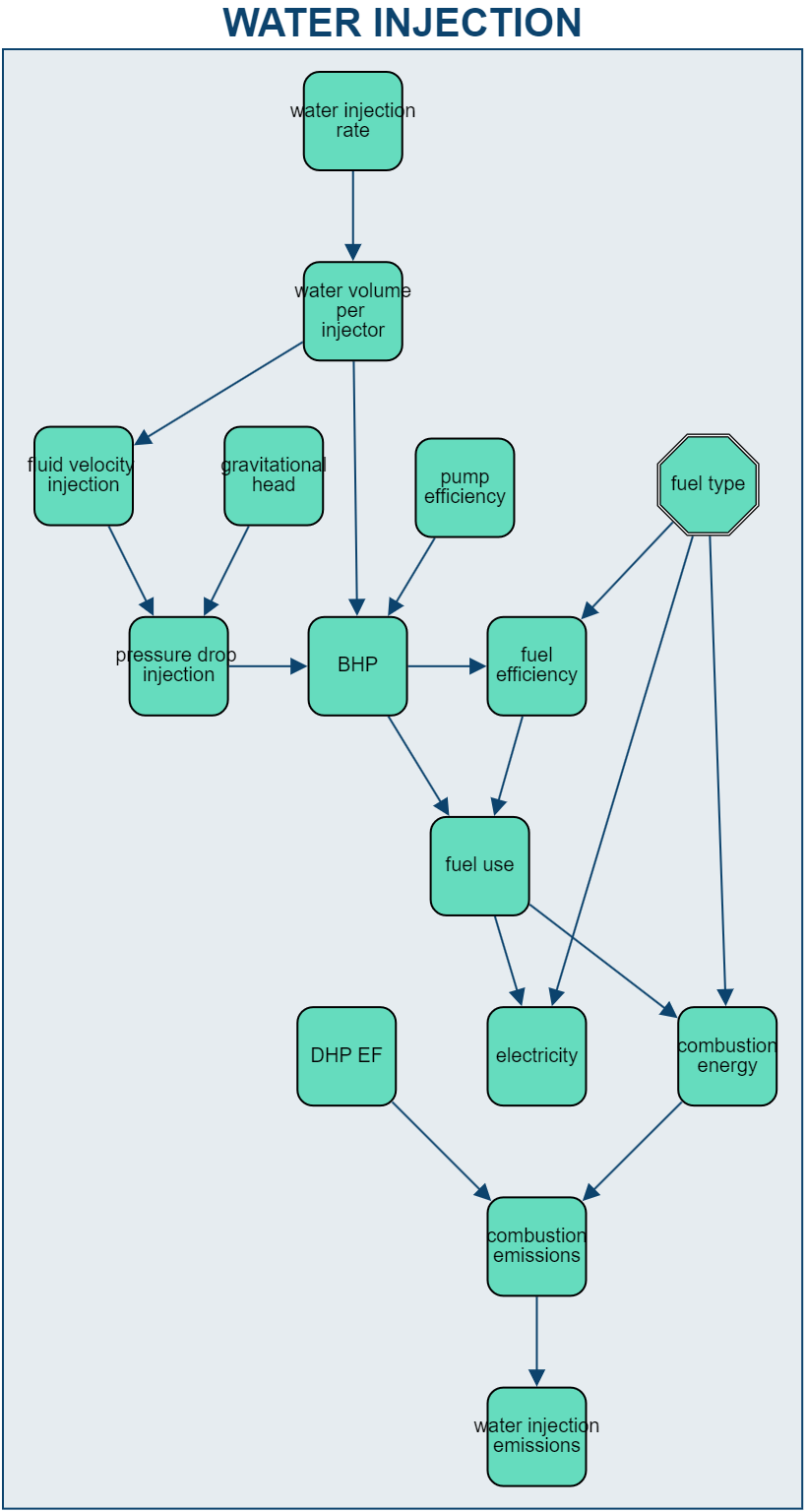}
    \caption{Water Injection module}
    \label{fig:ewater_injection}
\end{figure}

\subsection{Crude oil storage tank}

\begin{figure}[H]
    \centering
    \includegraphics[width=1\linewidth]{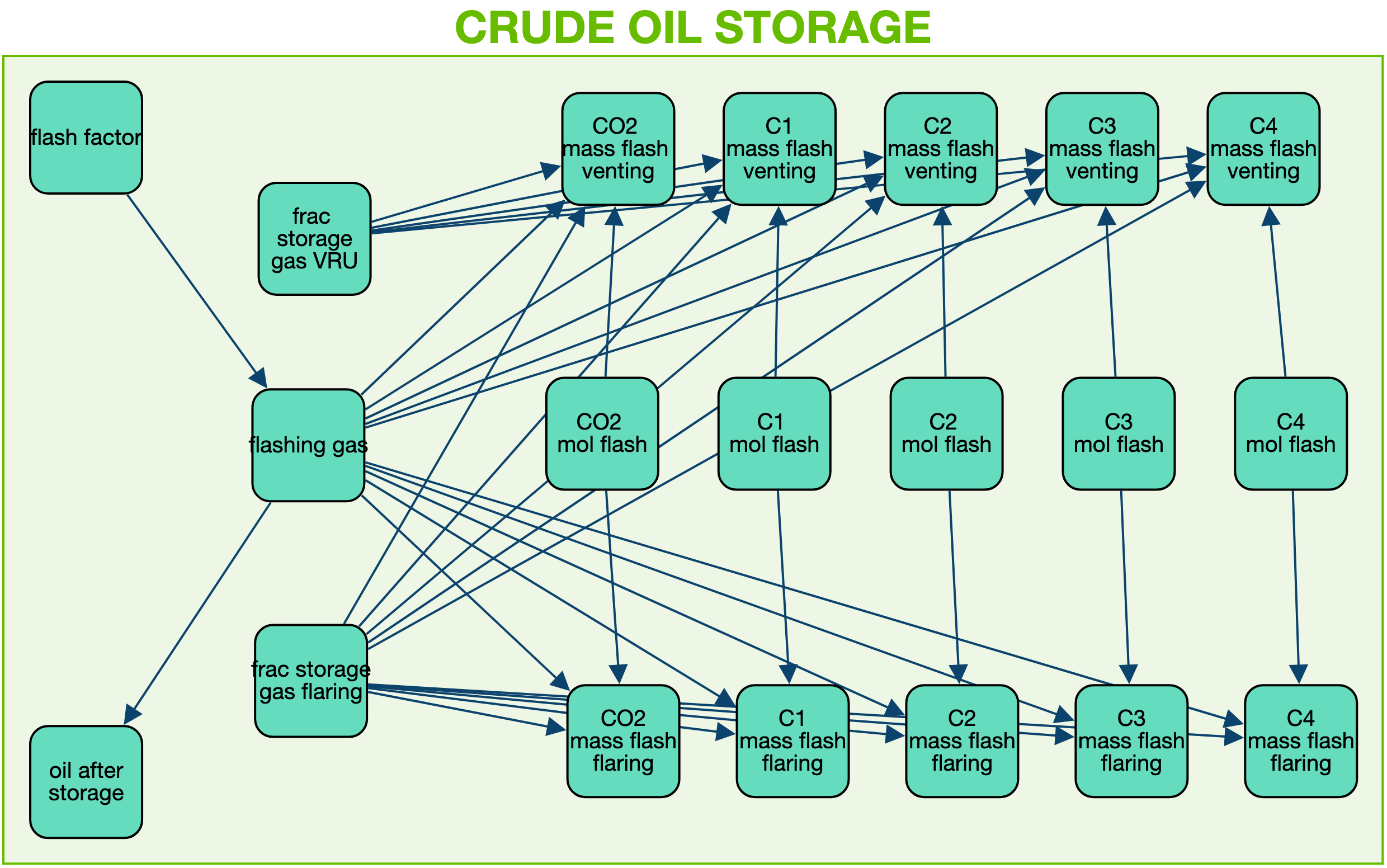}
    \caption{Crude Oil Storage Module}
    \label{fig:tank_storage}
\end{figure}

Crude oil storage tanks (Figure \ref{fig:tank_storage}) are a significant source of greenhouse gas (GHG) emissions in the oil and gas industry. Emissions occur primarily through loss of working and breathing, where volatile organic compounds (VOCs) evaporate from the oil. Flashing losses also contribute significantly; they occur when oil is pumped into or out of the tank, leading to a sudden change in pressure that causes some VOCs to evaporate. The degree of GHG emissions from oil tank storage varies depending on factors such as tank type, oil type, and operating conditions. The storage of oil tanks is estimated to account for approximately 10\% of the GHG emissions in the oil and gas industry.

During normal operations, the amount of vapor leaving the solution and entering the tank headspace - a process known as "flashing" - is a key driver of the emissions from the storage of crude oil. Flashing emissions are notably variable and represent a significant source of uncertainty, as evidenced by the wide ranges reported in empirical studies and various thermodynamic or correlational models. In OPGEE, empirical results from the HARC/TERC study on tank flashing emissions are utilized \cite{Hendler2009}.

\subsection{Global warming potential and emissions factor}

\begin{figure}[H]
    \centering
    \includegraphics[width=1\linewidth]{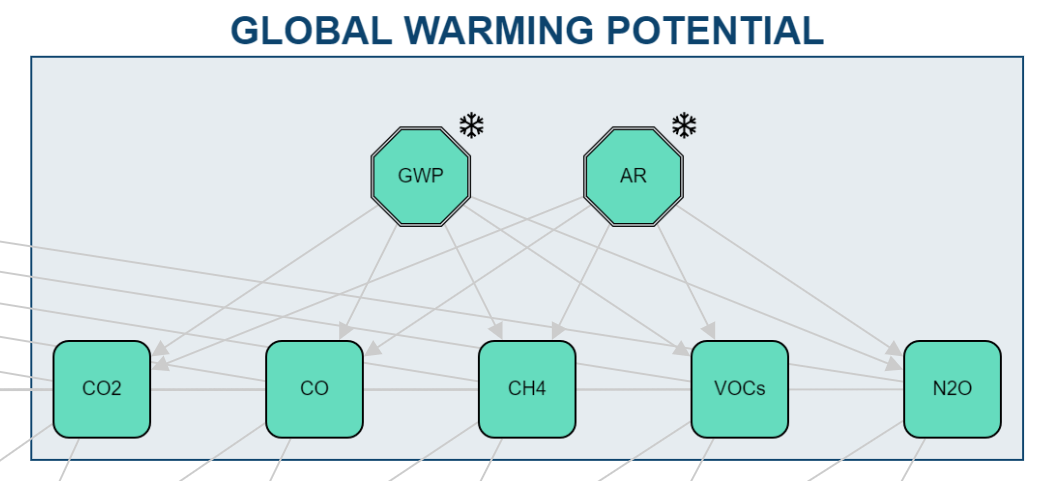}
    \caption{Global Warming Potential}
    \label{fig:gwp}
\end{figure}

Global Warming Potentials (GWPs) for gases with radiative forcing are presented with multiple options for consideration, as shown in Figure \ref{fig:gwp}. These options include values from the Intergovernmental Panel on Climate Change's (IPCC) Fourth Assessment Report \cite{IPCC2007}, and the Fifth Assessment Report (AR5) \cite{IPCC2013}, both inclusive and exclusive of climate carbon feedbacks. Users can select GWPs based on a 20-year or a 100-year time horizon.

Emission factors are crucial for calculating greenhouse gas (GHG) emissions from both combustion (fuel combustion) and non-combustion (venting and fugitives) sources. For fuel combustion, emission factors are obtained from CA-GREET \cite{Wang2009}. The range of gases tracked includes VOC, CO, CH$_{4}$, N$_{2}$O, and CO$_{2}$. These emissions are converted into carbon dioxide equivalents using the IPCC's Global Warming Potentials (GWPs) \cite{IPCC2007}. The conversion is carried out as follows:

\begin{equation}
    \label{eq:emissions_conversion}
    EM_{CO_{2}eq,i} = EM_{i} \cdot \text{GWP}_{i}
\end{equation}

\noindent where \( EM_{CO_{2}eq,i} \) denotes the emissions of species \( i \) in carbon dioxide equivalent [gCO$_{2}$eq], \( EM_{i} \) represents the emissions of species \( i \) [g], and GWP$_{i}$ is the GWP of species \( i \) [gCO$_2$eq/g].

\clearpage
\section{Benchmark: Global oilfields}
In recent years, several efforts have been made to establish a global benchmark for oilfield GHG emissions. Notable among these efforts are three publicly available databases, which publish emission metrics such as Carbon Intensity (CI) and emissions for an exhaustive list of over 50,000 oilfields worldwide. These databases, which include \href{https://www.archieinitiative.org/}{The Archie Initiative}, \href{https://fossilfuelregistry.org/}{Fossil Fuel Registry}, and \href{https://ociplus.rmi.org/}{OCI+}, base their emissions computations on open-source lifecycle assessments (LCA) such as OPGEE. Their approach has been validated by multiple studies published in reputable journals (\cite{Dixit2023}, \cite{Masnadi2021}, \cite{Liang2020}).

\subsection{Validation of the PGM Model}

\begin{figure}[H]
    \centering
    \includegraphics[width=1\linewidth]{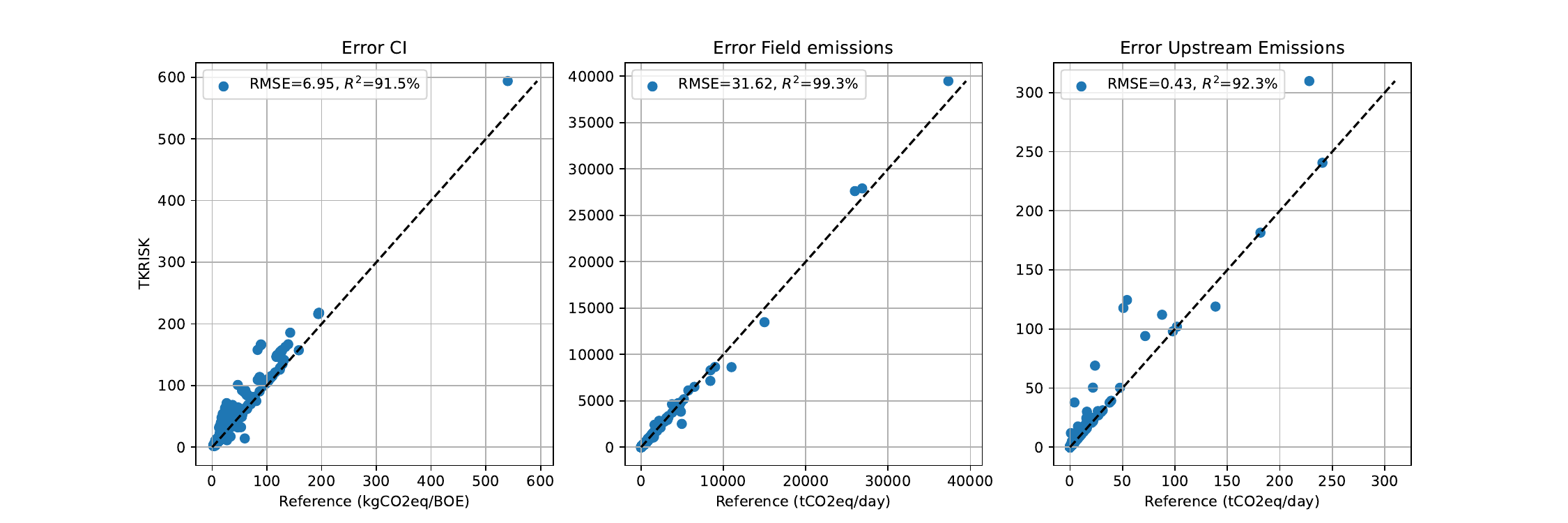}
    \caption{Cross plot comparison of CI (left), Field Emissions (middle), and upstream emissions (right) obtained from OPGEE v3.0c and the graph-based model (OPGM) for 1150 benchmark oilfields}
    \label{fig:global_oilfield_crossplot}
\end{figure}

We have conducted a systematic comparison of GHG analytics generated by our Oilfield Pollutant Graphical Model (OPGM) with those produced by the most recent reference model, OPGEE v3.0c, on a benchmark of global oilfields. The results of this comparison are illustrated in Figure~\ref{fig:global_oilfield_crossplot}.

The comparison indicates that all metrics align closely with those obtained from the reference model. To our knowledge, the PGM model can generate results that are as accurate as those of the latest version of OPGEE, v3.0c. As an example, the root mean square error (RMSE) on the average CI is 6.95 kgCO2eq / BOE. This error lies within the measurement error made on oifields using state-of-the-art sampling technologies (\cite{Rutherford2021}). The total run-time for analyzing 1150 fields using the OPGM model is less than 1 second, compared to over 3 hours using OPGEE v3.0c.

\section{Analysis}
\subsection{Life of field emissions}
We present an analysis of the emissions of an oilfield over its useful life (approximately 45 years). This highlights how different phases of operations lead to varying emissions patterns and intensities. Until then, most analysis using OPGEE focused on average emissions using a single time-step approach. OPGM allows us to seamlessly perform time-dependent estimations. To illustrate, we use a synthetic field whose characteristics are presented below.

\begin{table}[H]
    \centering
    \caption{Reservoir properties}
    \label{tab:oilfield characteriristcs}
    \begin{tabular}{rc}
        \toprule
        Property& Value\\
        \midrule
         Location& Offshore\\
         Fluid& Oil\\
         Reserves& 490 MMB\\
         Reservoir Depth& 3400 ft\\
         Reservoir Temperature& 131 F\\
         API& 18$^o$\\
         Number of Producers& 5\\
         Number of Injectors (W/G)& 3/2\\
         Plateau& 70 MBPD\\
         GOR& 1370 stb/scf\\
         FOR& 100-900\\
         Venting fraction& 0.002\\
         Flaring combution efficiency& 93.2\%\\
         Water injection rate/well& 15 MBPD\\
         \bottomrule
    \end{tabular}
\end{table}

\begin{figure}[H]
    \centering
    \includegraphics[width=1\linewidth]{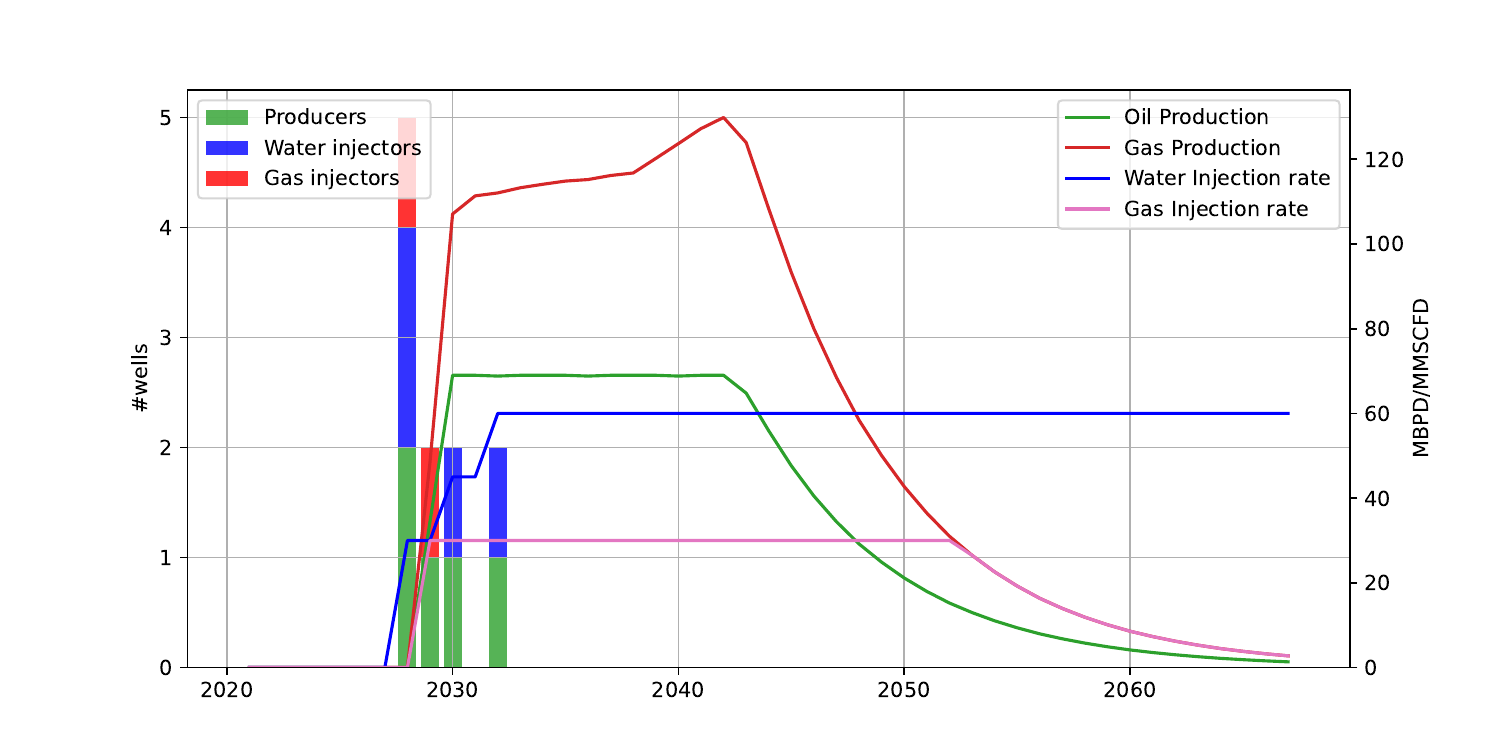}
    \caption{Oilfield production profile (measured on the right axis) and well count (measured on left axis)}
    \label{fig:production_profile}
\end{figure}

The field production profile is presented in Figure~\ref{fig:production_profile}

\begin{figure}[H]
    \centering
    \includegraphics[width=1\linewidth]{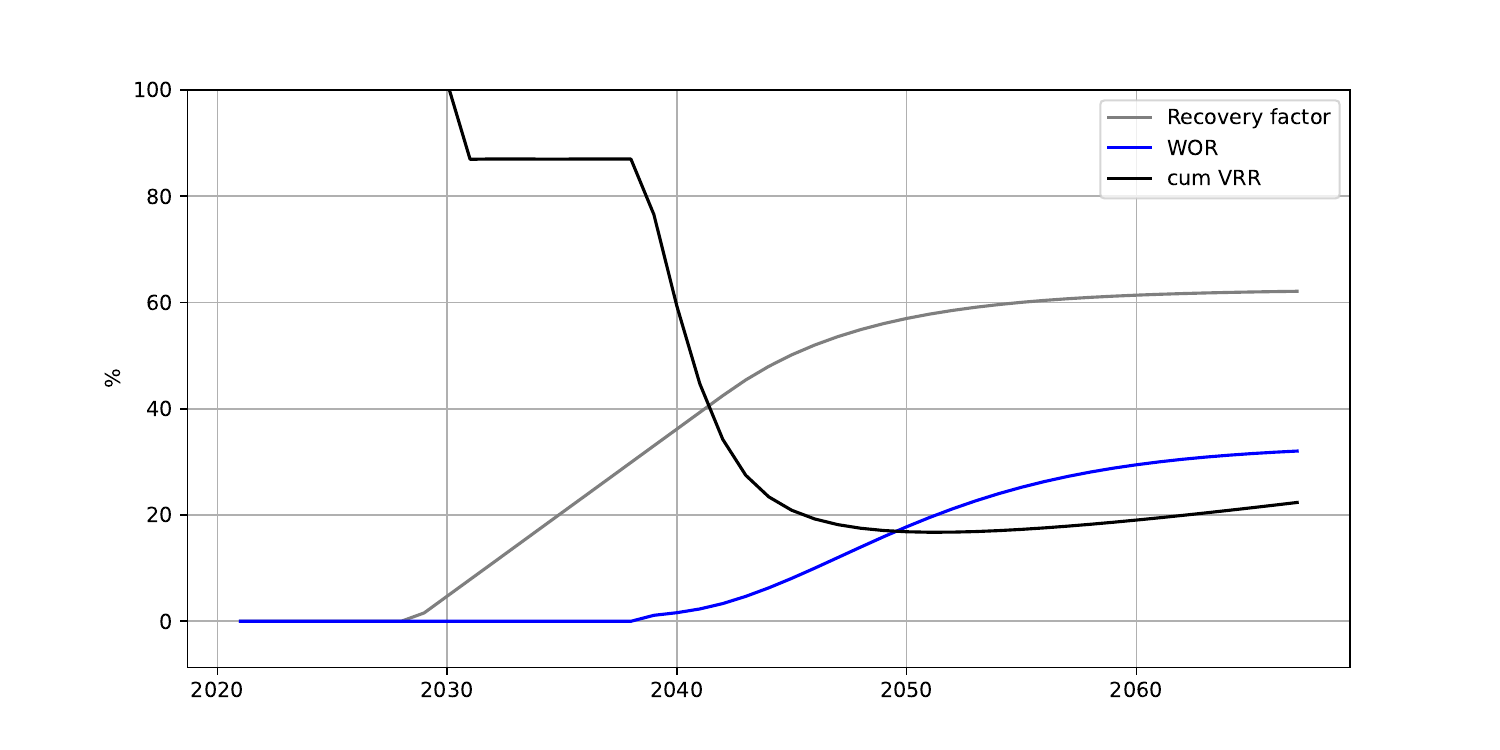}
    \caption{Production ratios, recovery factor in grey, voidage replacement ratio in black and Water Oil ratio (WOR) in blue.}
    \label{fig:ratios_profile}
\end{figure}

\begin{figure}[H]
    \centering
    \includegraphics[width=1\linewidth]{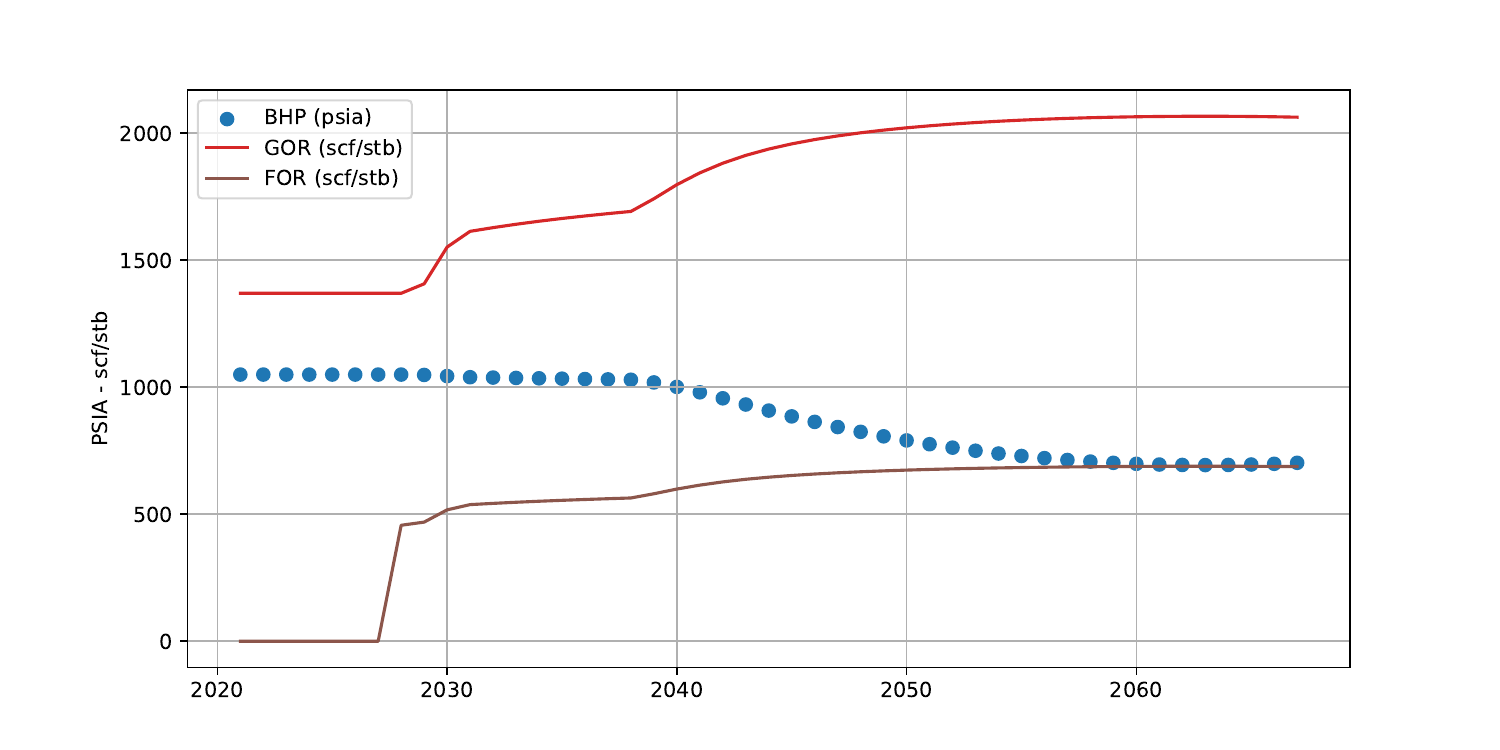}
    \caption{Pressure and gas to oil ratio in the field}
    \label{fig:pressure_profile}
\end{figure}

This profile shows a plateau production that lasts 13 years. Note the increasing gas production during the plateau years due to the depressurization of the reservoir. We also assume that water production increases with time as shown in Figure~\ref{fig:ratios_profile}. 
Both the water and the gas produced are re-injected into the reservoir for disposal and as a pressure maintenance strategy keeping a Voidage Replacement Ratio (VRR) between 50\% and 75\% for most of the recovery period. This model accounts for the gas separation within the reservoir under the effect of depletion. Figure~\ref{fig:pressure_profile} shows the joint evolutions of the GOR and the bottom hole pressure. 
The fugitive to oil ratio (FOR) is a constant fraction of the GOR.
 
\begin{figure}[H]
    \centering
    \includegraphics[width=1\linewidth]{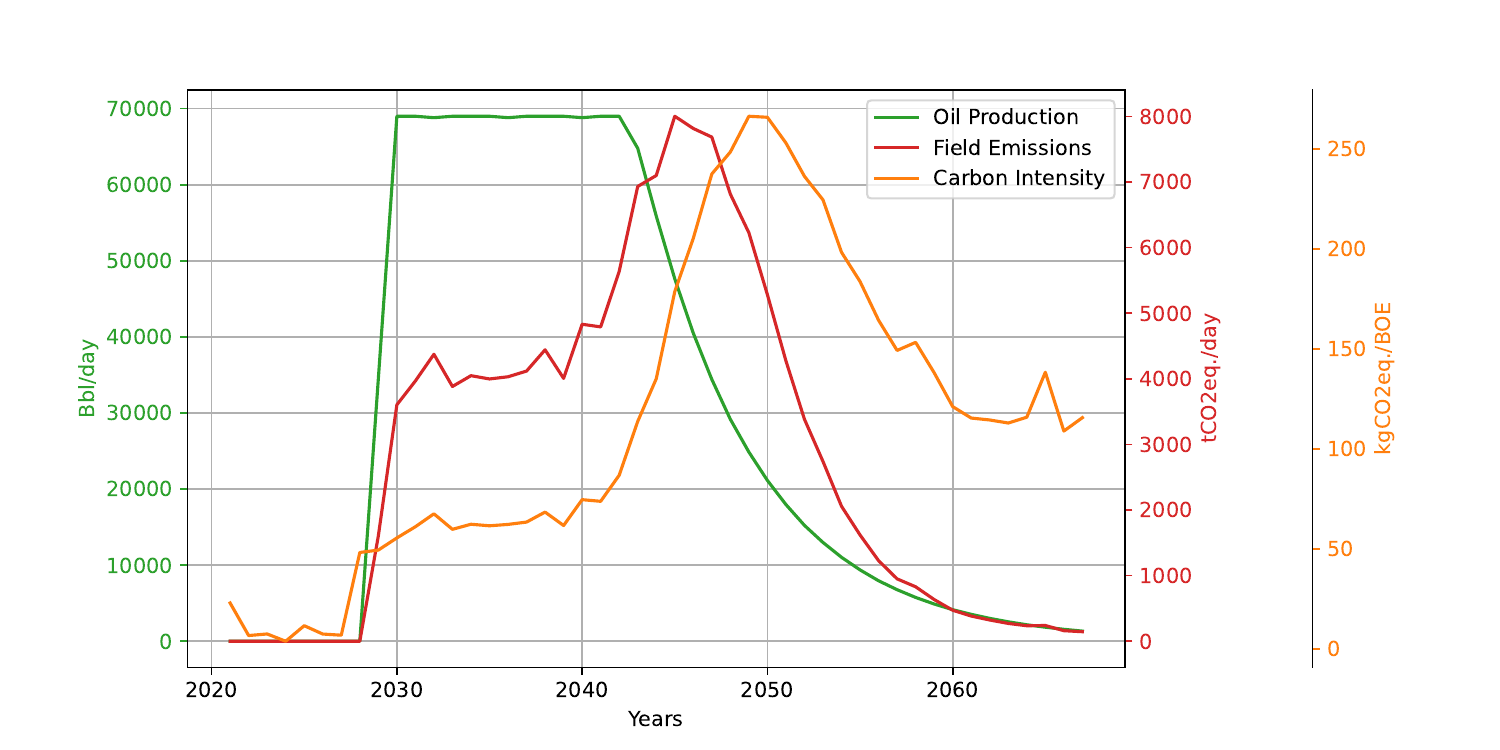}
    \caption{Comparative evolution of oil production (green), field emissions (red) and carbon intensity (orange)}
    \label{fig:emission_profile}
\end{figure}

We run OPGM on the full production profile and generate emissions for each time step. Figure~\ref{fig:emission_profile} shows how the emission profile evolves with production.
We note that the field's emissions increases towards the end of the plateau life. This seemingly counter-intuitive result can be explained by the large volume of water produced toward the end of the field's life. Indeed, we can breakdown the field's total emissions in seven categories:
\begin{enumerate}
    \item Downhole pump
    \item Separation 
    \item Flaring
    \item Venting
    \item Water injection
    \item Vapor Recovery Unit (VRU)
    \item Upstream
\end{enumerate}
Figure~\ref{fig:emissions_breakdown} shows the evolution of each categories, in absolute and relative to the total field's emissions:
\begin{figure}[H]
    \centering
    \includegraphics[width=1\linewidth]{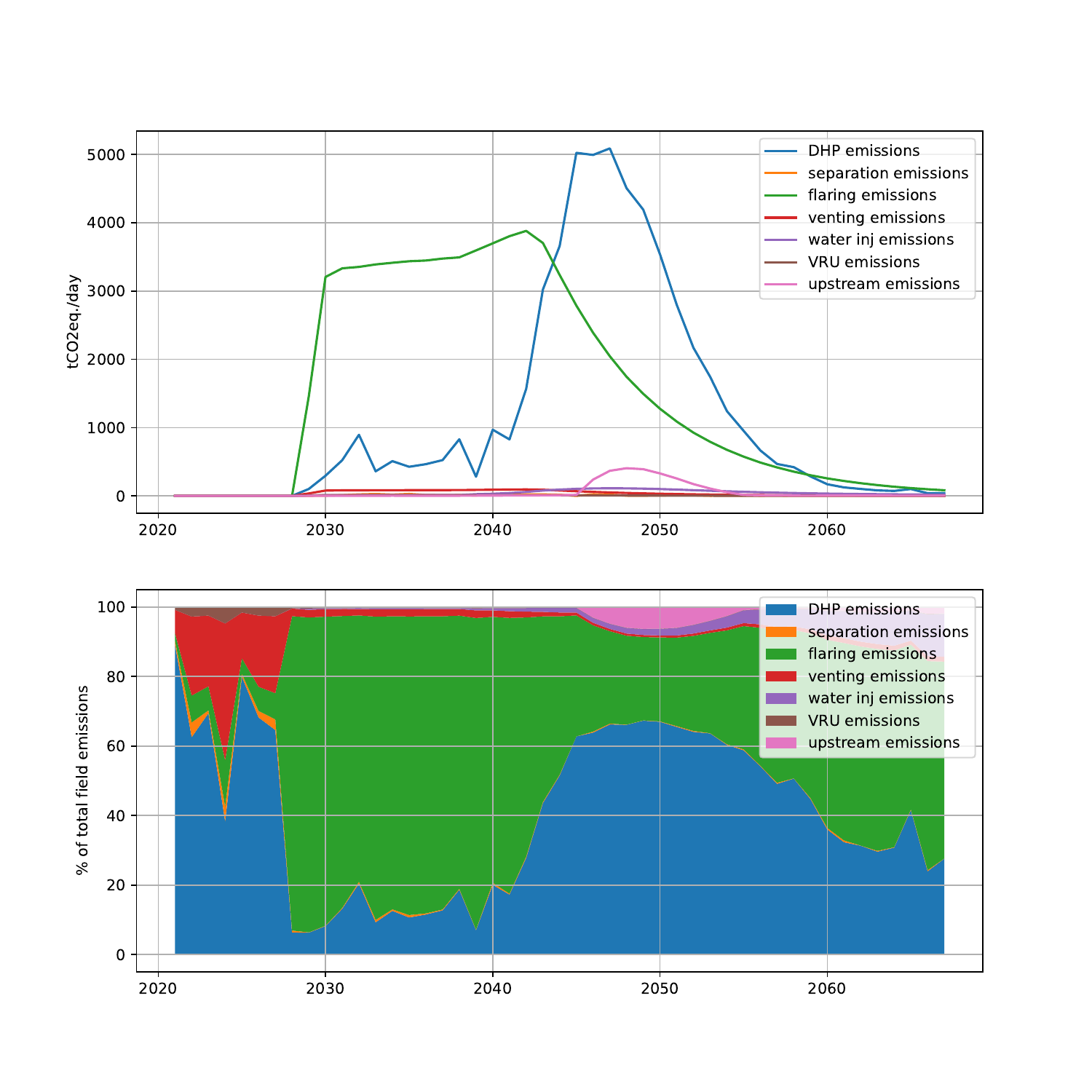}
    \caption{Field's emissions breakdown over time represented in absolute (top) and relative to total field's emissions (bottom)}
    \label{fig:emissions_breakdown}
\end{figure}
We note that while flaring is the main contributor while gas production remains high, the downhole pump emissions become overwhelming as the water production increases and the field needs an increasing amount of energy to lift the heavy fluid column out of the reservoir. OPGM offers the possibility to present an even more granular breakdown, as each node in the graphical model can be "exposed" to the user. This helps diagnose the main contributors to the emissions and act appropriately to alleviate them.

\subsection{Country-level CI}
\begin{figure}
    \centering
    \includegraphics[width=0.5\linewidth]{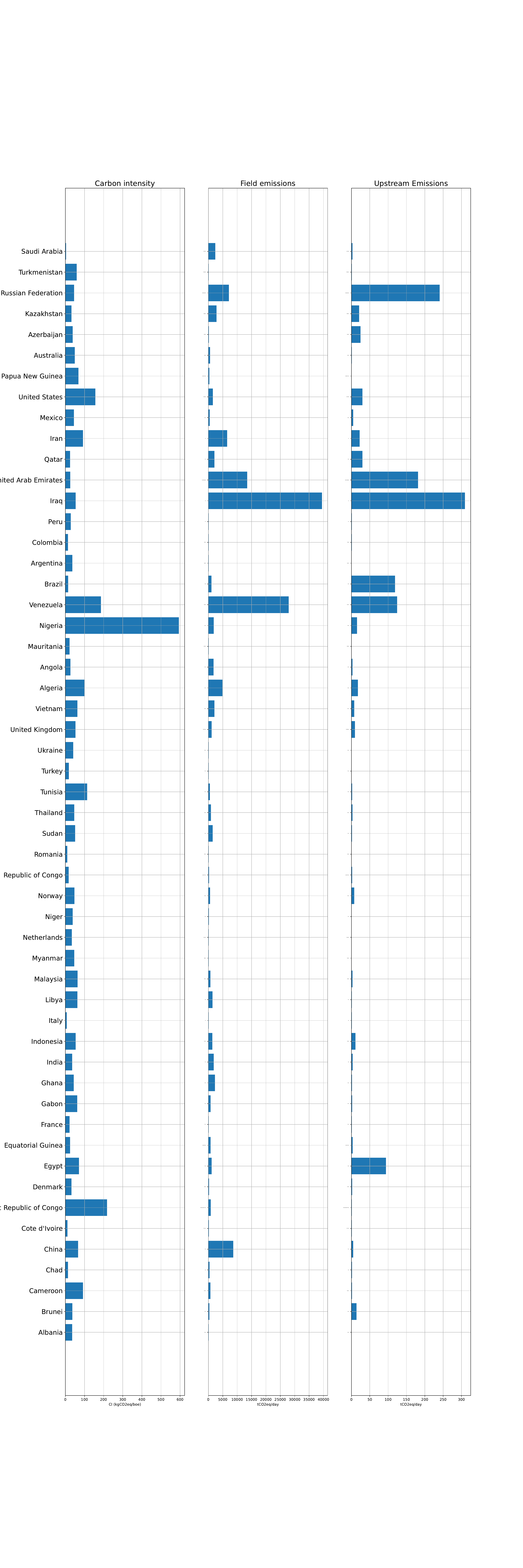}
    \caption{Average emissions indicators per field per country. Carbon intensity (left), Field Emissions (middle), Upstream emissions (right)}
    \label{fig:global_oilfields_countries}
\end{figure}

Besed on the 1150 fields considered in our global benchmark, we compute some key metrics rolled up at the country level. We represent rolled averages (per field within a country) as this helps to emphasize areas where remediation could have the largest impact per unit of production. Results are represented in Figure~\ref{fig:global_oilfields_countries}

\clearpage

\subsection{Uncertainty analysis}
OPGM is designed to be probabilistic by construct. This feature allows for a natural analysis of the impact of uncertainties. By converting one or more nodes in the graph from deterministic to probabilistic with a range of possible outcomes, users can observe how uncertainties propagate through the graph, influencing various nodes, including the output metrics. TKRISK\textsuperscript{\textregistered} improves this capability by offering simulations of a wide array of continuous and discrete distributions, parameterized according to the standards set by open source statistics libraries such as SciPy (\cite{2020SciPy-NMeth}).

To illustrate this analysis, we present a series of examples:

\begin{enumerate}
    \item South Pokamasovskoye oilfield in Russia:
    
    \begin{figure}[H]
        \centering
        \includegraphics[width=0.8\linewidth]{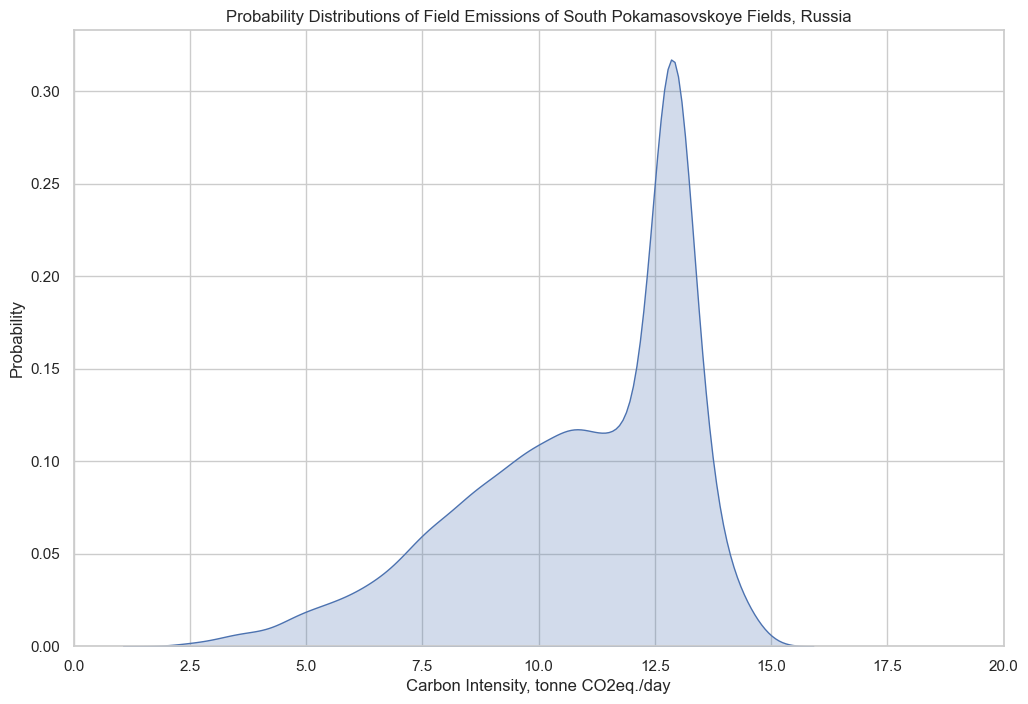}
        \caption{Probability distribution of the carbon intensity for the South Pokamasovskoye field in Russia.}
        \label{fig:uncertainty_field_CI}
    \end{figure}

    \item A comprehensive analysis of 215 fields in Russia:
    
    \begin{figure}[H]
        \centering
        \includegraphics[width=0.8\linewidth]{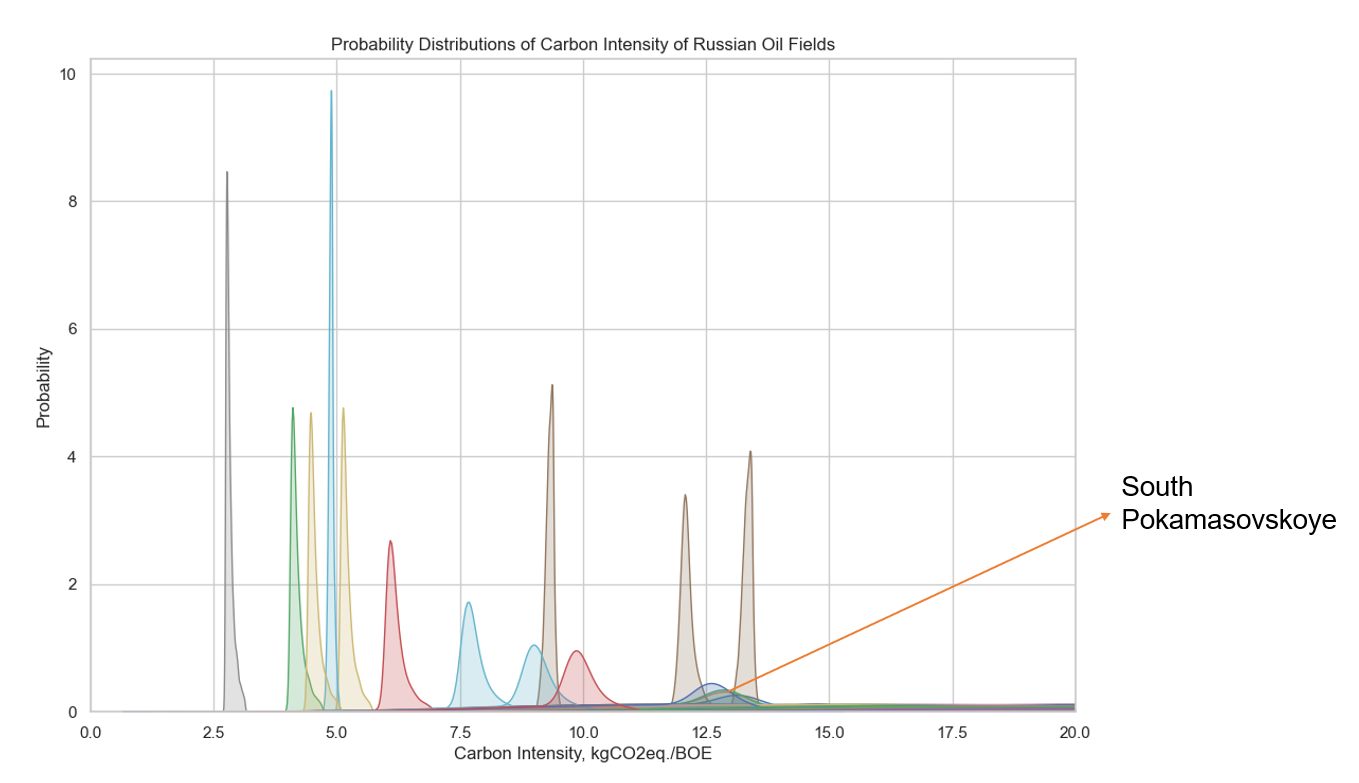}
        \caption{Probability distribution of carbon intensity across Russian oil fields.}
        \label{fig:uncertainty_country_CI}
    \end{figure}

    \item A comparison at the country level, demonstrating how the model aggregates data to generate a national volume-weighted average of crude oil upstream GHG intensities. The global volume-weighted carbon CI estimate is also presented, showing the range from the 5th to 95th percentiles of the Monte Carlo simulation, which addresses the uncertainty associated with missing input data.
    
    \begin{figure}[H]
        \centering
        \includegraphics[width=1\linewidth]{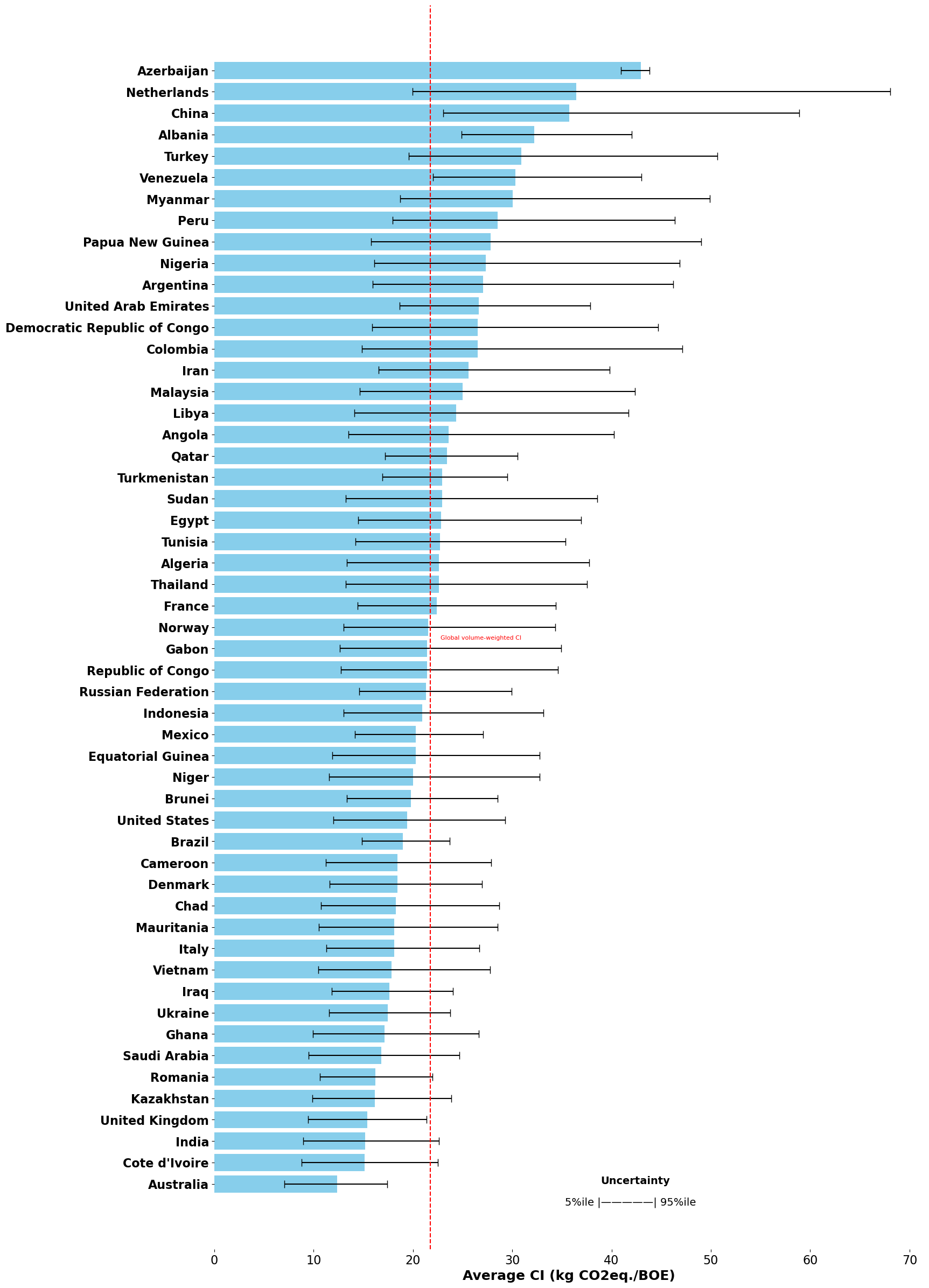}
        \caption{National volume-weighted-average crude oil upstream GHG intensities. The global volume-weighted carbon CI estimate is shown (red line, 21.8 kg CO2eq./BOE), with error bars reflecting the 5th to 95th percentiles of the Monte Carlo simulation.}
        \label{fig:Country_average_CI}
    \end{figure}
\end{enumerate}
The amortized run-time for analyzing one field using the PGM model is less than 1e-5 seconds, compared to over 10 seconds using OPGEE v3.0c.

The detailed uncertainty analysis can be found in the Appendix \ref{sec:uncertainty_analysis}.

\section{Discussion}

The integration of the Oilfield Pollutant Graphical Model (OPGM) into the established framework of the Oil Production Greenhouse Gas Emission Estimator (OPGEE) represents a paradigm shift in emissions benchmarking and uncertainty analysis in global oilfield operations. This analysis underscores the critical contributions of the OPGM, delineating its role in advancing the state-of-the-art in lifecycle assessment (LCA) tools.

OPGM notably maintains the essential elements of its predecessor, OPGEE v3.0c, ensuring that it retains the same level of thoroughness and rigor. This commitment to preserving established standards of accuracy and reliability is crucial for the model's continued relevance and trustworthiness in environmental impact assessments.

A defining attribute of OPGM is its user-focused design, which simplifies complex emission estimation processes. This design enhances user interaction, making the model more accessible and transparent, thereby improving the interpretability of its analysis. OPGM's ability to visually track data through each computational step increases user understanding and confidence in the model's results.

The introduction of probabilistic graphical modeling techniques within the OPGM framework marks a significant methodological advancement. This feature allows for an extensive exploration of emissions scenarios, capturing the inherent variability and uncertainty prevalent in environmental data. The model's flexibility in transitioning between deterministic and probabilistic nodes endows it with a unique ability to conduct sensitivity analyses, thereby providing a comprehensive understanding of the influence of various input parameters on the outcome metrics.

Perhaps most notably, OPGM demonstrates extraordinary computational efficiency, a quantum leap from the traditional Excel-based OPGEE model. This efficiency not only underscores the model's ability to handle extensive datasets with ease but also marks it as a tool eminently suitable for large-scale, comprehensive environmental inventories and benchmarking exercises.

In summary, OPGM emerges as an intellectually robust, user-friendly, and computationally agile enhancement to the LCA toolkit. It represents a significant leap forward in emissions analysis, marrying accuracy with efficiency. As the oil and gas industry grapples with the urgent need for precise and scalable environmental assessments, the OPGM stands as a beacon of innovation, poised to redefine standard practices. Looking ahead, the potential integration of the OPGM into broader environmental and policy frameworks heralds a new era of dynamic, data-driven decision-making, pivotal for steering the industry toward more sustainable horizons.

% \section{Nomenclature}

% \begin{itemize}
%     \item \( B_{ob} \): The bubble-point oil formation volume factor represents the expansion of a barrel of oil at the bubblepoint pressure in the reservoir compared to its volume at stock tank conditions.
%     \item \( a_1, a_2, a_3, a_4, a_5 \): Empirical constants specific to the correlation, determined through experimental data.
%     \item \( R_s \): The solution gas-oil ratio, indicating the amount of gas dissolved in the oil at a specific reservoir pressure and temperature.
%     \item \( T_r \): The reservoir temperature, usually in degrees Fahrenheit (°F).
%     \item \( \gamma_o \): Oil gravity, measured in degrees API, indicates the density of the oil relative to water.
%     \item \( \gamma_g \): Gas gravity measures the density of the gas relative to air or a standard such as pure methane.
% \end{itemize}

% \clearpage

\clearpage
\appendix
\section{Formation volume factor}
The formation volume factor is a critical parameter in petroleum engineering, which defines the changes in the volume of hydrocarbons as they are produced from the reservoir to the surface conditions. This factor is essential for oil and gas, as it significantly varies depending on phase behavior, temperature, and pressure conditions.

The oil formation volume factor, \( B_o \), is the ratio of the volume that a barrel of oil occupies in the reservoir to its volume under stock tank (standard) conditions. It is a function of the solution gas-oil ratio, temperature, and pressure (the gas in solution being the largest determinant). The oil FVF at bubble point pressure is generated using bubble point oil FVF correlations, with specific references to the work of Al-Shammasi \cite[Table A-3, p. 317]{Fanchi2007}\cite{Alshammasi2001}:

\begin{equation}
B_{ob} = a_1+ a_2 R_s (T_r - 60) + a_3\left( \frac{R_s }{\gamma_o}\right) + a_4  \left(\frac{T_r - 60}{\gamma_o}\right)+ a_5 R_s \left( \frac{\gamma_g}{\gamma_o}\right)
\end{equation}

The isothermal compressibility of oil, \( c_o \), used in the above equation, is derived from the correlation by Al-Marhoun \cite{Almarhoun1992}.

\begin{equation}
c_o = 55.233 \times 10^{-6} - 60.588\times 10^{-6}\gamma_o
\end{equation}

Similarly, the gas formation volume factor is the ratio of the volume of gas at the temperature and pressure of the stream to the volume of the same amount (moles) of gas under standard conditions. The gas volume ratio is calculated as follows:

\begin{equation}
FVF_g= \frac{Z p_{0} T}{p T_{0}}
\end{equation}

Where \( Z \) is the Z-factor that describes the deviation of a real gas from the behavior of the ideal gas. This formula incorporates the absolute temperature of the stream \( T \), ambient absolute temperature \( T_0 \), pressure of the stream \( p \), and ambient pressure \( p_0 \).

\section{Thermodynamic properties and compressor}
\begin{figure}[H]
    \centering
    \includegraphics[width=1\linewidth]{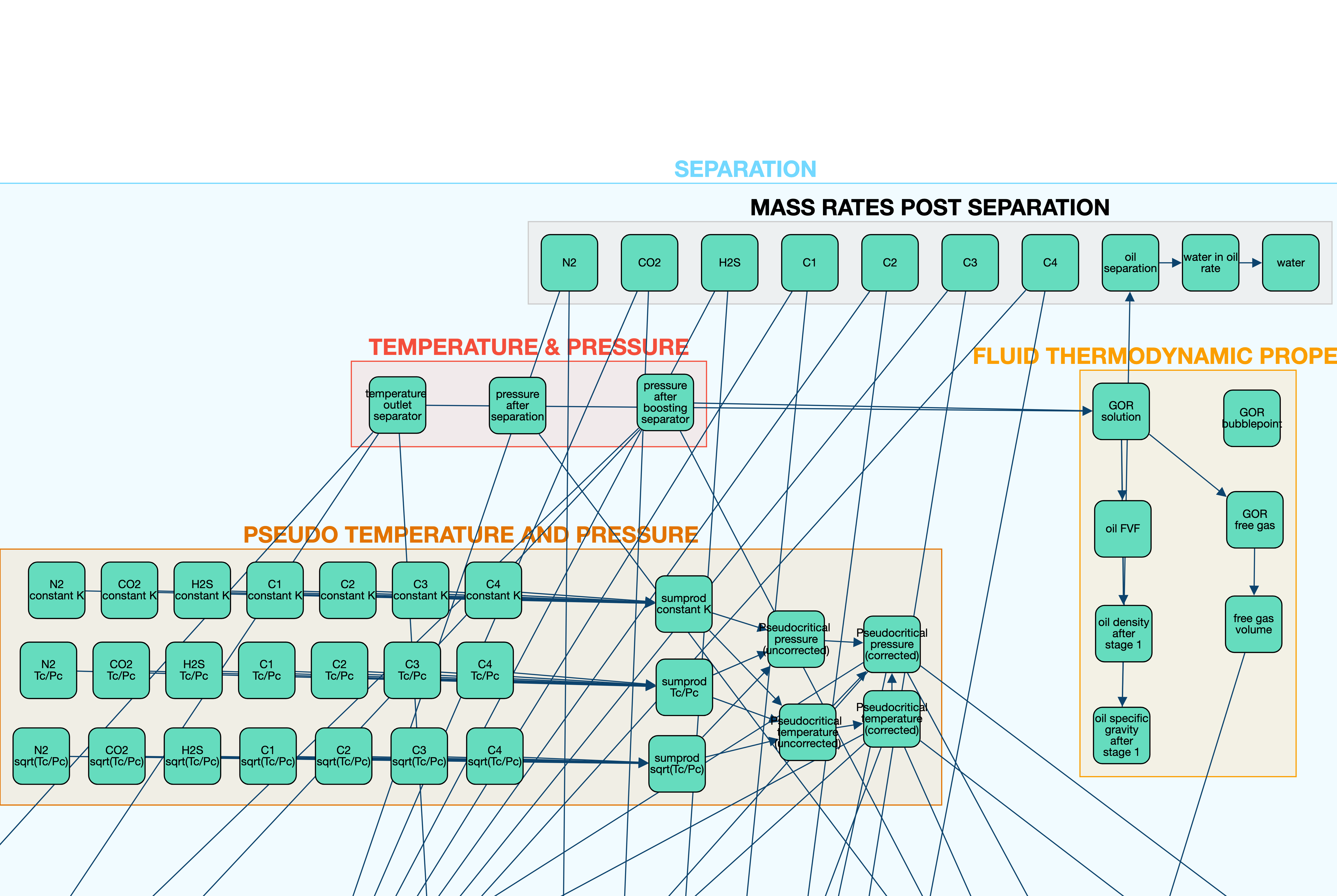}
    \caption{Thermodynamics group in Separation module}
    \label{fig:separation_thermodynamics}
\end{figure}

Separation, as it involves the phase changes of multicomponent mixtures, involves complex thermodynamic interactions. The model encompasses a comprehensive treatment of the thermodynamics as shown in figure~\ref{fig:separation_thermodynamics}
The model implements a rigorous treatment of mass flow rates between its components. This was a major change between OPGEE v2 and OPGEE v3, and we follow the latest available implementation.

In petroleum engineering, the dynamics of gas compression play a crucial role in the efficient handling and processing of natural gas. This section discusses calculating the total gas compression ratio, the stages of multistage compression, and the workings of gas reinjection compressors.

The total gas compression ratio is calculated using the following:
\begin{equation} \label{eq:compression_ratio}
R_{C} = \frac{p_{d}}{p_{s}}
\end{equation}
where \( p_{d} \) is the discharge pressure [psia], and \( p_{s} \) is the suction pressure [psia].

Multi-stage compression is applicable when the compression ratio \( R_C \) exceeds five. This is crucial because compressors can only handle limited temperature changes and multiple stages allow cooling between stages, making compression more efficient and closer to an isothermal process.

\begin{equation} \label{eq:compression_stages}
\text{If} \,\,\frac{p_{d}}{p_{s}}\, \textless \, 5, \,\, \text{then}\,\, R_C=\frac{p_{d}}{p_{s}},\,\, \text{otherwise if}\,\,\left(\frac{p_{d}}{p_{s}}\right)^{\frac{1}{N}}\, < \,5,\,\,\text{then}\,\, R_C=\left(\frac{p_{d}}{p_{s}}\right)^{\frac{1}{N}}
\end{equation}
where \( N \) is the number of compression stages, and \( p_{d} \) and \( p_{s} \) are as previously defined.

\begin{figure}[H]
    \centering
    \includegraphics[width=1\linewidth]{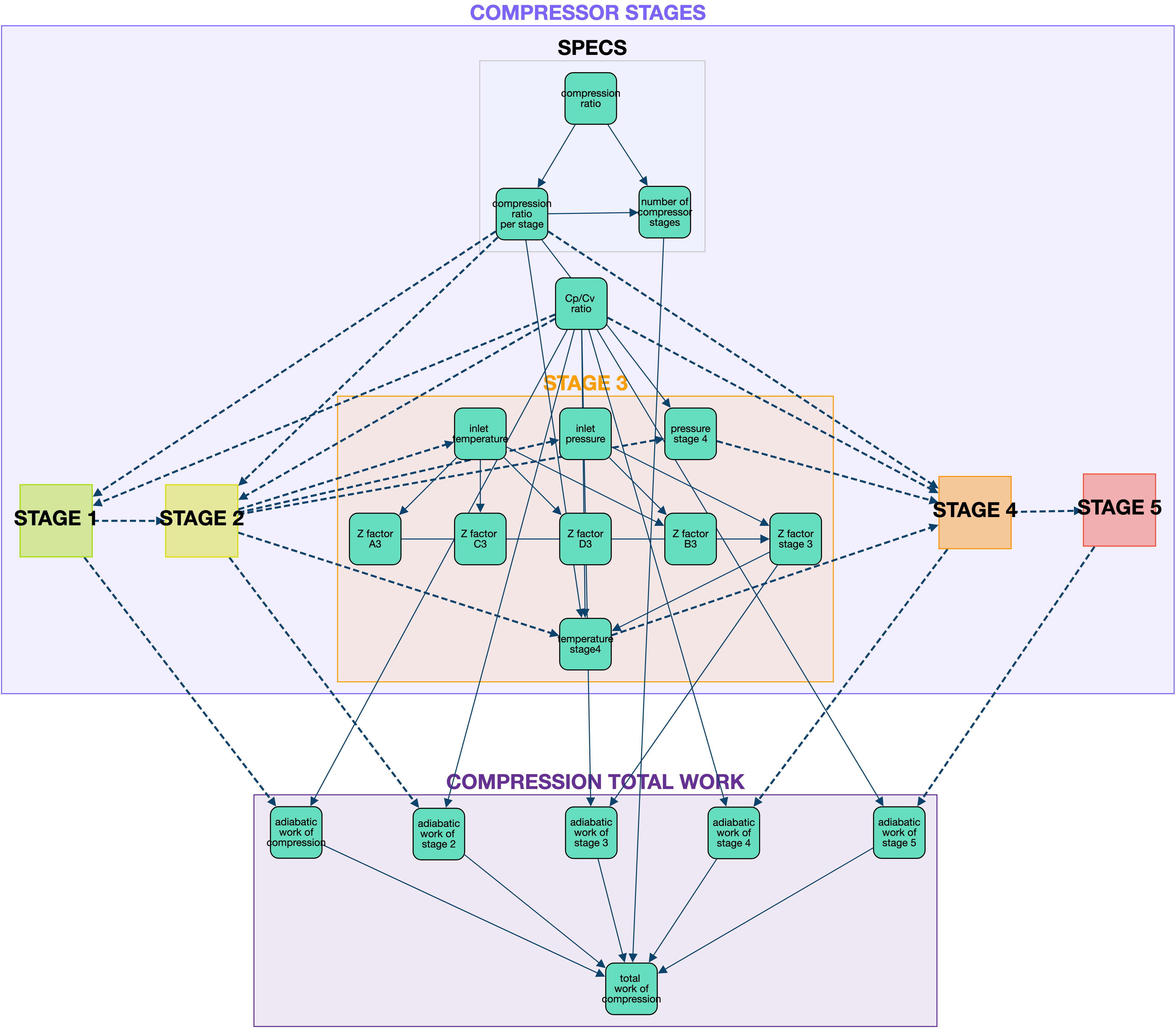}
    \caption{Compressor groups within the separation module}
    \label{fig:sep-compressor_stages}
\end{figure}

Here, we present an example with five stages of compression. This number can be arbitrarily increased by copying \& pasting the existing stages, as shown in Figure~\ref{fig:sep-compressor_stages}.

\section{Flaring efficiency}
Carbon dioxide equivalent emissions from flaring are calculated based on the flaring volume and flare efficiency. Flare efficiency, the fraction of flared gas combusted, varies with factors such as flare exit velocities and diameters, crosswind speeds, and gas composition \cite{Johnson2001, Johnson2008}. In Alberta, for example, flare efficiencies have been estimated to range from 55\% to \( \geq 99\% \), with a median value of 95\%, adjusted for wind speed distributions \cite{Johnson2008}.

Due to limited data availability in many cases, OPGEE utilizes empirical flare efficiency measurements from several studies \cite{Ozumba2000, Gvakharia2017, Chambers2003, Zavala2021, Caulton2014a}. These studies, which include 79 measurements from locations including the United States-Bakken, Nigeria, Mexico, Canada, and the United States-Pennsylvania, offer insights into the efficiency of methane flare destruction. 

\begin{table}[h]
\centering
\begin{scriptsize}
\caption{Empirical Measurements of Methane Flare Destruction Efficiency}
\label{tab:empirical_flare_eff}
\begin{tabular}{@{}ccccccccc@{}}
\toprule
Study & Location & Year & $N$ obs. & Mean & Median & Min & Max & SD \\ \midrule
Caulton & USA & 2014 & 13 & 99.915 & 99.953 & 99.717 & 99.983 & 0.079 \\
Chambers & Canada & 2003 & 5 & 89.000 & 91.000 & 74.000 & 98.000 & 8.983 \\
Gvakharia & USA & 2017 & 52 & 95.496 & 96.956 & 80.189 & 99.966 & 4.603 \\
Ozumba & Nigeria & 1998 & 8 & 98.488 & 99.000 & 95.700 & 99.200 & 1.189 \\
Zavala-Araiza & Mexico & 2021 & 1 & 94.000 & 94.000 & 94.000 & 94.000 & - \\ \midrule
All & Var. & Var. & 79 & 96.108 & 97.732 & 74.000 & 99.983 & 4.948 \\ \bottomrule
\end{tabular}
\end{scriptsize}
\end{table}

For deterministic analysis, the mean flare efficiency from all observations in Table \ref{tab:empirical_flare_eff} is used, which is 96.108\%. In uncertainty analysis, each draw is based on a random selection of the measurements in the table.

\section{Water treatment table}
\label{sec:water_treatment_table}

\begin{table}
\begin{scriptsize}
\caption{Categorization of water treatment technologies. Table based on table from Vlasopoulos et al. \cite{Vlasopoulos2006}, with minor modifications.}
\label{tab:stages}
\begin{tabular*}{0.75\columnwidth}{p{0.5\columnwidth}p{0.2\columnwidth}}
\toprule
Name & Signifier \\
\midrule
Stage 1 & \\
\midrule
Dissolved air flotation & DAF \\
Hydrocyclones & HYDRO \\ 
& \\
Stage 2 & \\
\midrule
Rotating biological contactors & RBC \\
Absorbents & ABS \\ 
Activated sludge & AS \\
Trickling filters & TF \\
Air stripping & AIR \\
Aerated lagoons & AL \\
Wetlands & CWL \\ 
Microfiltration & MF \\
& \\
Stage 3 & \\
\midrule
Dual media filtration & DMF \\ 
Granular activated carbon & GAC \\
Slow sand filtration & SSF \\
Ozone & OZO \\ 
Organoclay & ORG \\ 
Ultrafiltration & UF\\ 
Nanofiltration & NF \\ 
& \\
Stage 4 & \\ 
\midrule
Reverse osmosis & RO \\ 
Electrodialysis reversal & EDR \\ 
Warm lime softening & WLS \\
\bottomrule
\end{tabular*}
\end{scriptsize}
\end{table}

\section{Fuel consumption}
The energy consumption calculated from each process must be converted into fuel use by looking up the fuel use factor of the primary mover. OPGEE provides four types of drivers for pumps, compressors, and onsite electricity generators. Drivers include natural gas-powered engines, gas turbines, diesel engines, and electric motors. Table \ref{tab:drivers_size} shows the types and size ranges of the drivers included in the model.

\begin{table}
\begin{scriptsize}
\caption{Types and size ranges of the drivers embedded in OPGEE.}
\label{tab:drivers_size}
\begin{tabular*}{\columnwidth}{@{\extracolsep{\fill}}p{0.25\columnwidth}p{0.15\columnwidth}p{0.20\columnwidth}p{0.35\columnwidth}}
\toprule
Type & Fuel & Size range [bhp] & Reference \\
\midrule
Internal combustion engine & Natural gas & 95 - 2,744 &  \cite{Caterpillar2012} \\
Internal combustion engine & Diesel & 1590 - 20,500 &  \cite{Caterpillar2012} \\
Simple turbine & Natural gas & 384 - 2,792 &   \cite{Solarturbines2012}\\
Motor & Electricity & 1.47 - 804 & \cite{GE2011}\\
\bottomrule
\end{tabular*}
\end{scriptsize}
\end{table}

\begin{figure}
\centering
\begin{subfigure}{0.4\textwidth}
    \includegraphics[width=\textwidth]{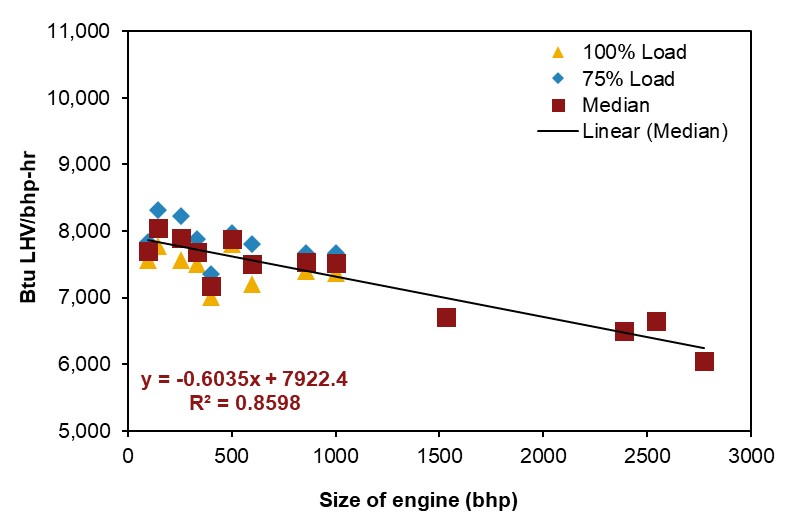}
    \caption{Natural gas engine efficiency under 100 \%, 75 \%, and median load of power}
\end{subfigure}
\hfill
\begin{subfigure}{0.4\textwidth}
    \includegraphics[width=\textwidth]{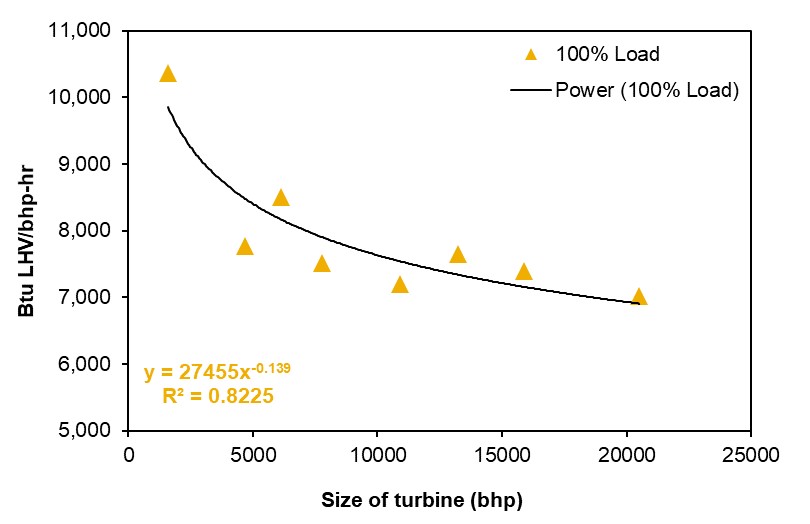}
    \caption{Natural gas turbine efficiency under 100 \% load of power}
\end{subfigure}
\hfill
\begin{subfigure}{0.4\textwidth}
    \includegraphics[width=\textwidth]{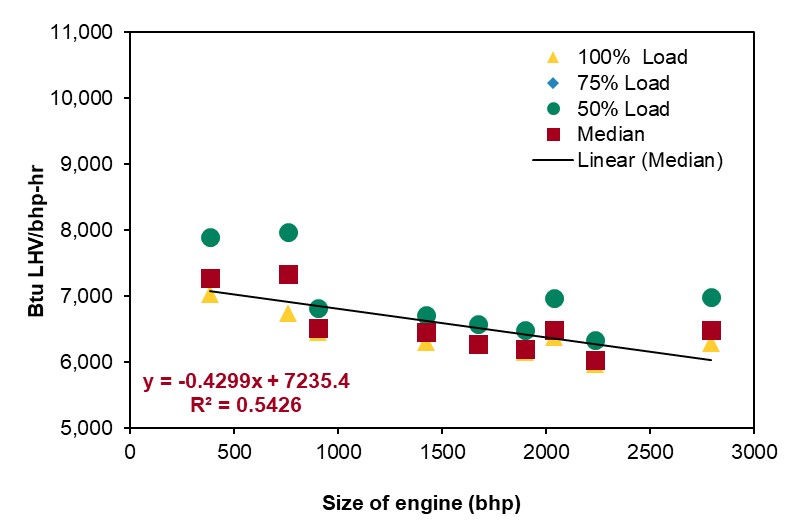}
    \caption{Diesel engine efficiency under 100 \%, 75 \%, 50 \%, and median load of power}
\end{subfigure}
\hfill
\begin{subfigure}{0.4\textwidth}
    \includegraphics[width=\textwidth]{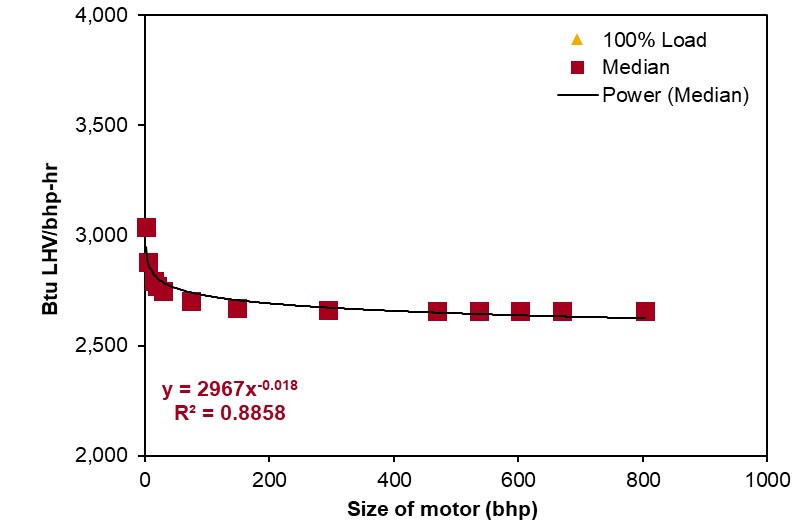}
    \caption{Electric motor efficiency under 100 \% and median load of power}
\end{subfigure}
        
\caption{Correlation between the size of the engine (BHP) and efficiency (Btu/BHP-hr) from (a) Natural gas engine, (b) Natural gas turbine, (c) Diesel engine, and (d) Electric motor.}
\label{fig:driver_correlation}
\end{figure}

In the process, the primary mover type is chosen from either the default of the user or the model. Given the type of primary mover and the horsepower of the brake, the efficiency of the primary mover is calculated using the correlations in Figure \ref{fig:driver_correlation}. The figure collects data from the reference in Table \ref{tab:drivers_size}, including the size, efficiency, and power load of the engine from the various types of real-world drivers to generate the correlation. The correlation between engine size and efficiency is generated using the median load for natural gas engines, diesel engines, and electric motors and 100\% load for natural gas turbines (the data we collected is 100\% load). The correlation expressions are in the left-hand corner of each subfigure in Figure \ref{fig:driver_correlation}. The R-squares of the correlations are 0.86, 0.82, 0.54, and 0.85, respectively. Fuel consumption of the selected primary mover type can be calculated in each process using these expressions.

In particular, the model assumes 100\% use of the natural gas produced for the combustion of natural gas on site. Table \ref{tab:combustion_EF} lists the combustion technologies and fuels included in the model:

\begin{table}[h]
    \centering
    \begin{scriptsize}
    \caption{Combustion Technologies and Fuels}
    \label{tab:combustion_EF}
    \begin{tabular*}{\columnwidth}{@{\extracolsep{\fill}}lcccccc}
        \toprule
        & Natural gas & Diesel & Crude & Residual oil & Pet. coke & Coal \\
        \midrule
        Industrial boiler & \checkmark & \checkmark & \checkmark & \checkmark & \checkmark & \checkmark \\
        Turbine & \checkmark & \checkmark & & & & \\
        CC gas turbine & \checkmark & & & & & \\
        Reciprocating engine & \checkmark & \checkmark & & & & \\
        \bottomrule
    \end{tabular*}
    \end{scriptsize}
\end{table}

\section{Uncertainty analysis}
\label{sec:uncertainty_analysis}

The assessment of surface operations within the life-cycle assessment framework incorporates a variety of variables, each characterized by inherent uncertainties. To account for this, the model integrates a probabilistic approach, applying distinct statistical distributions to each parameter. This section delineates the variables pertinent to surface operations and their respective distributions.

\textbf{Flaring to Oil Ratio (FOR):} This variable captures the ratio of the volume of gas flared to the volume of oil produced, typically measured in standard cubic feet per barrel of oil (scf/bbl oil). The FOR is modeled as a normal distribution with a mean (\textit{loc}) of 107.1 and a standard deviation (\textit{scale}) of 35.7, bounded between a lower bound (\textit{lbound}) of 11 and an upper bound (\textit{ubound}) of 3010.
\begin{figure}[H]
    \centering
    \includegraphics[width=0.5\linewidth]{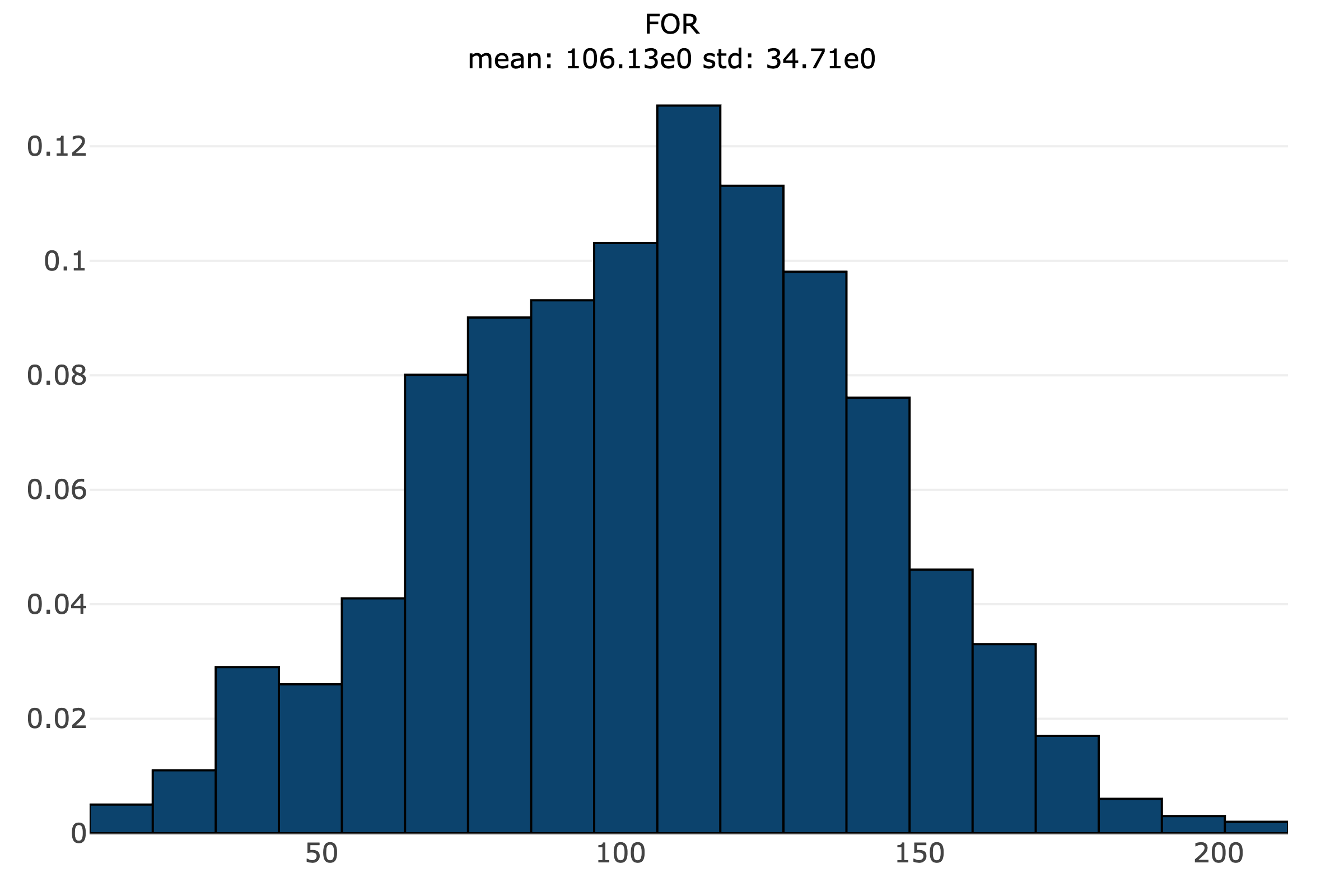}
    \caption{Flaring to Oil Ratio distribution}
    \label{fig:FOR}
\end{figure}

\textbf{Venting Fraction (venting\_frac):} This parameter represents the fraction of gas purposefully vented post-flaring. It adheres to a lognormal distribution, reflecting the multiplicative nature of the factors influencing this variable. The mean (\textit{loc}) is set at 0.002 with a standard deviation (\textit{scale}) of 0.001, bounded between 0 and 0.1.
\begin{figure}[H]
    \centering
    \includegraphics[width=0.5\linewidth]{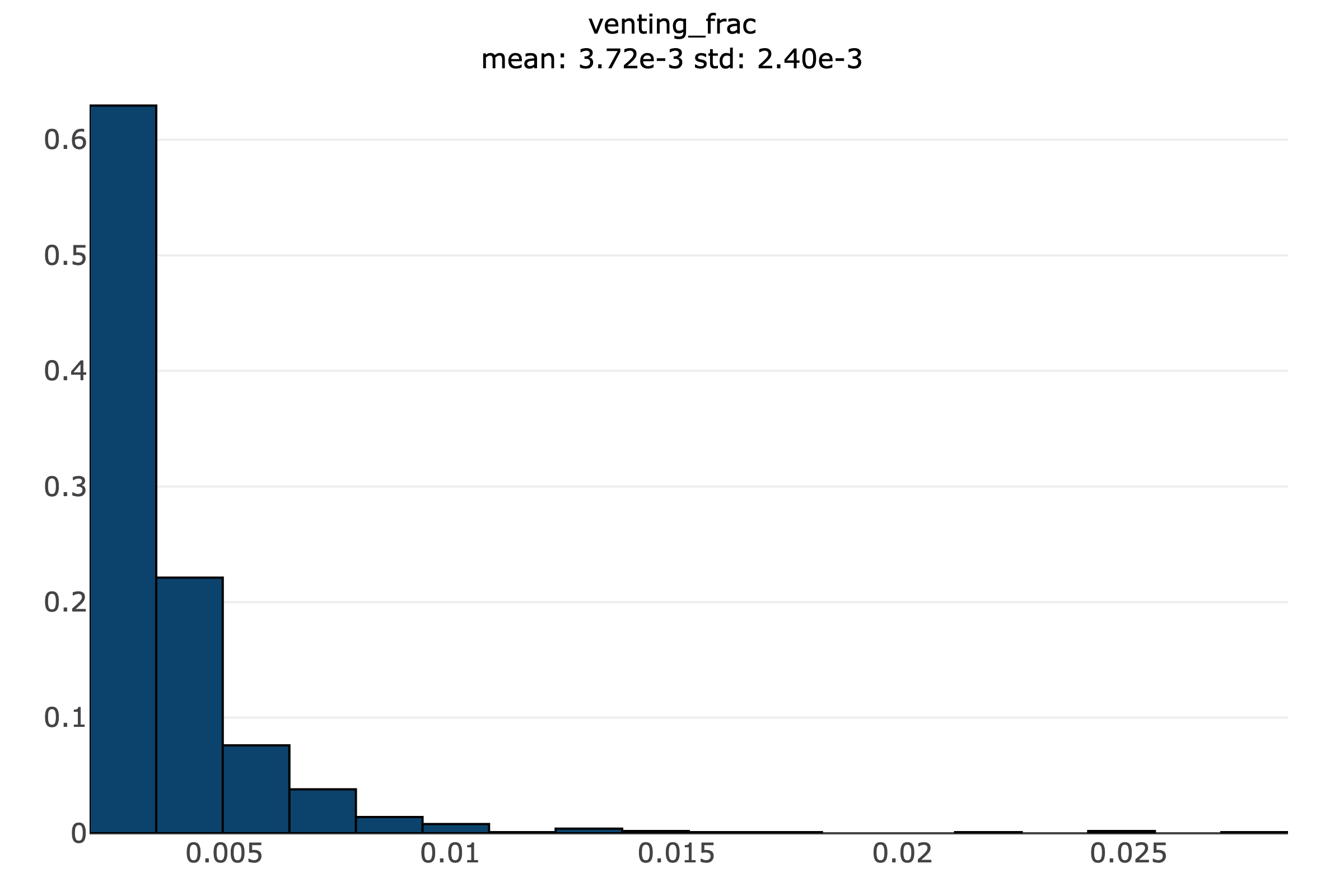}
    \caption{Venting Fraction distribution}
    \label{fig:venting_frac}
\end{figure}

\textbf{Fraction of Renewable Electricity (frac\_renewable\_elec):} This variable indicates the fraction of electricity generated onsite from renewable sources. A normal distribution models the variable with a mean (\textit{loc}) of 0.1 and a standard deviation (\textit{scale}) of 0.25, with bounds extending from 0 (no renewable electricity generation) to 1 (entirely renewable electricity generation).
\begin{figure}[H]
    \centering
    \includegraphics[width=0.5\linewidth]{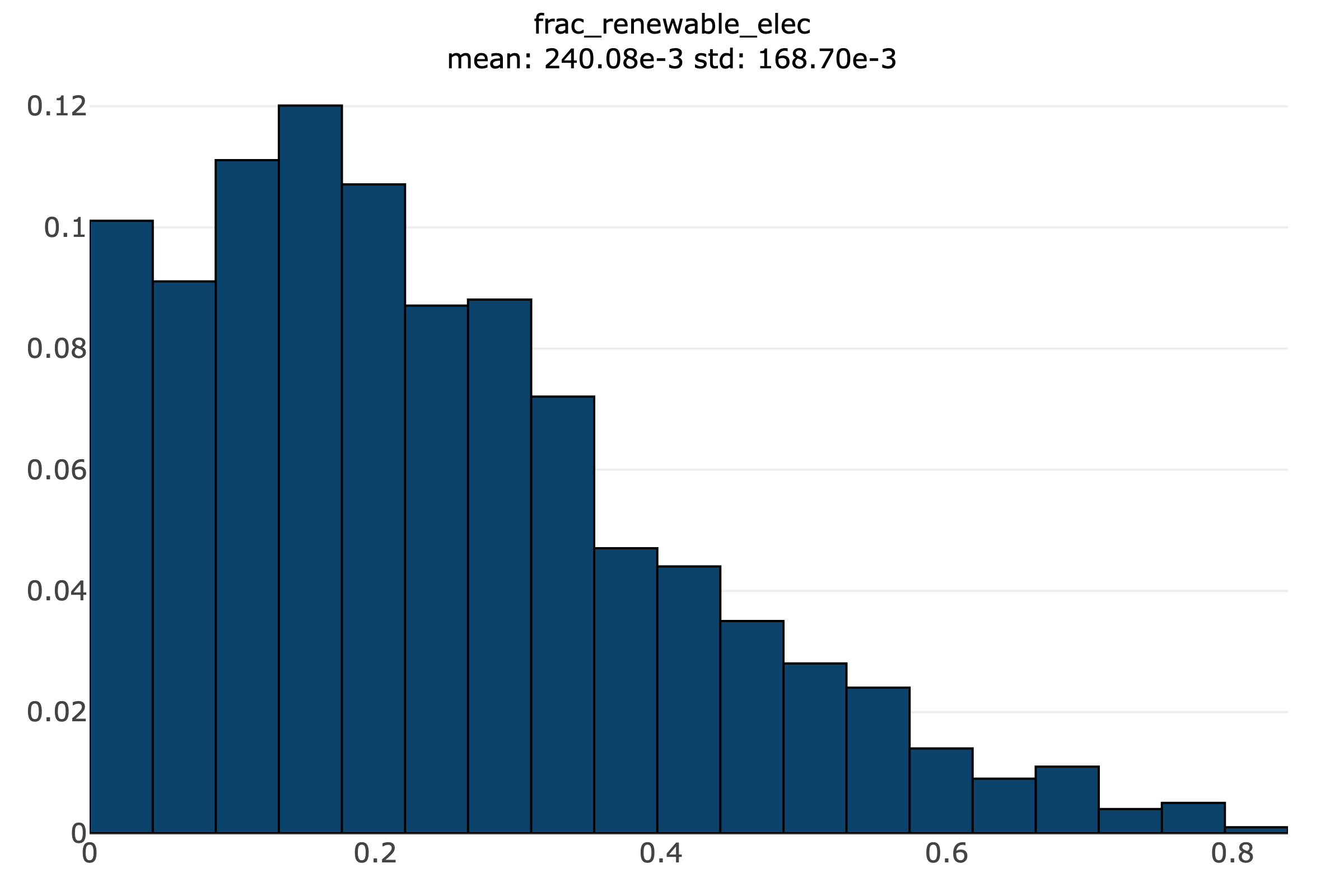}
    \caption{Fraction of Renewable Electricity distribution}
    \label{fig:frac_renewable_elec}
\end{figure}

\textbf{Separation Fugitive Loss Rate (separation\_loss\_rate):} The fugitive loss rate during the separation process is critical in assessing emissions. It is modeled using a uniform distribution to encapsulate the constant probability of loss rates within the specified range. The bounds are set from 0 to 0.00054.
\begin{figure}[H]
    \centering
    \includegraphics[width=0.5\linewidth]{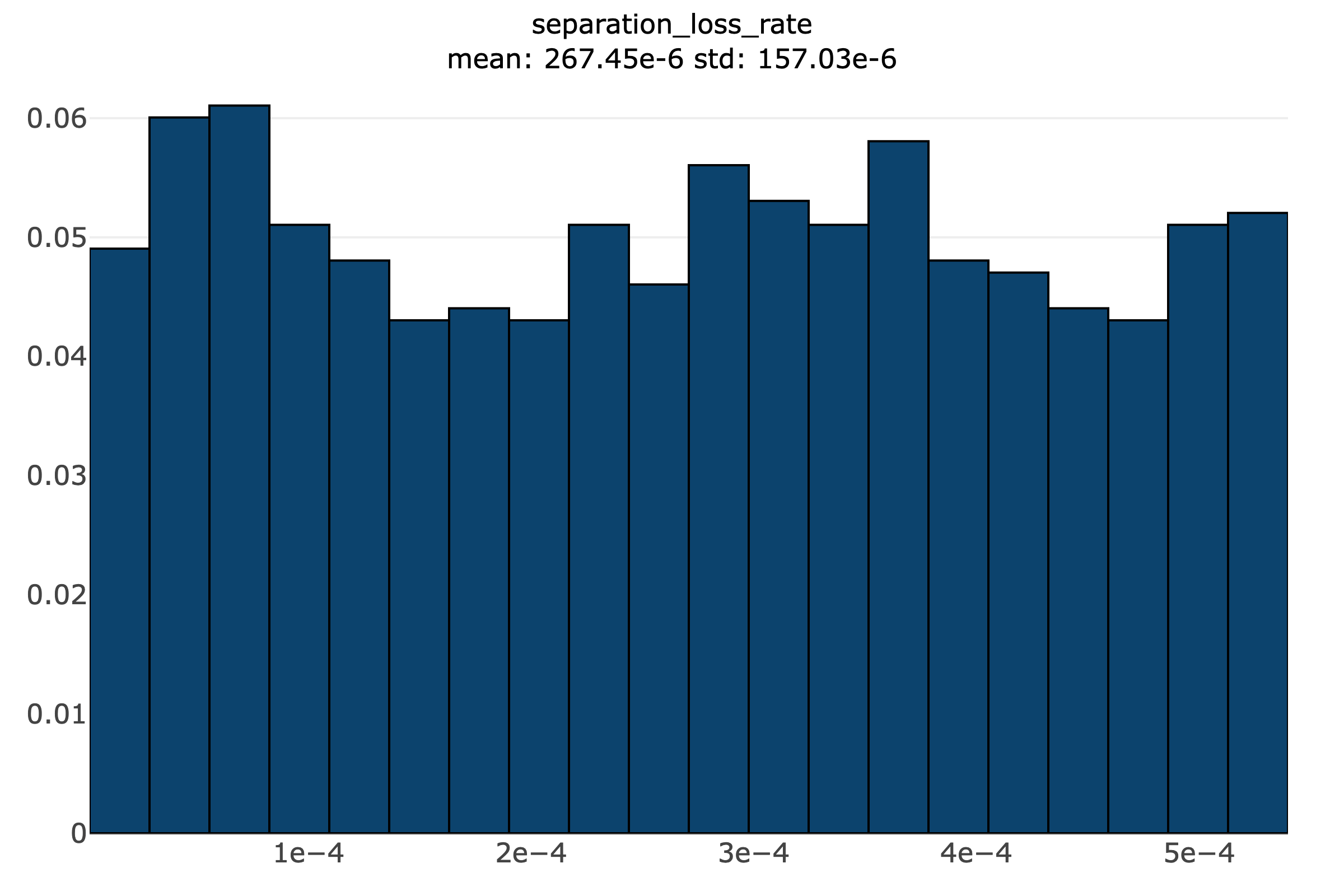}
    \caption{Separation Fugitive Loss Rate distribution}
    \label{fig:separation_loss_rate}
\end{figure}

\textbf{Downhole Pump Loss Rate (DHP\_loss\_rate):} The loss rate associated with downhole pumps is modeled with a lognormal distribution. The median value (\textit{loc}) is 0.000137, with a standard deviation (\textit{scale}) of 0.005096. The distribution is bounded by a lower limit of 0.001009 and an upper limit of 0.029909.
\begin{figure}[H]
    \centering
    \includegraphics[width=0.5\linewidth]{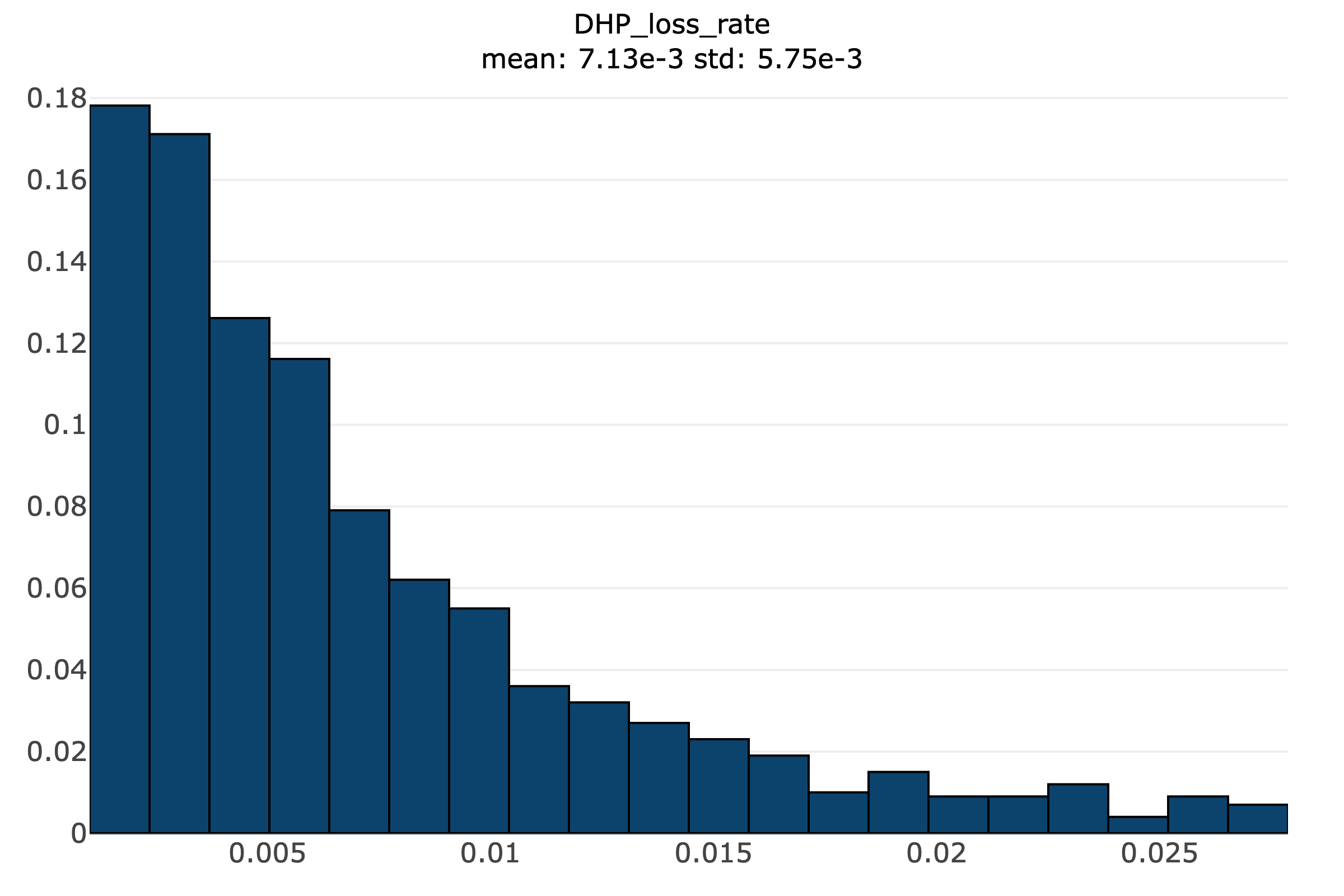}
    \caption{Downhole pump loss rate distribution}
    \label{fig:DHP_loss_rate}
\end{figure}

The distribution of each variable is chosen to best reflect the uncertainty and variability inherent in the respective surface operation processes.

\clearpage

\bibliography{main}

\begin{thebibliography}{10}

\bibitem{iea2023emissions}
{\sc I.~E. Agency}, {\em Emissions from oil and gas operations in net zero transitions}, 2023.

\bibitem{Almarhoun1992}
{\sc M.~A. Al-Marhoun et~al.}, {\em New correlation for formation volume factor of oil and gas mixtures}, Journal of Canadian Petroleum Technology, 31 (1992), pp.~22--26.

\bibitem{Alshammasi2001}
{\sc A.~Al-Shammasi}, {\em A review of bubblepoint pressure and oil formation volume factor corrrelations}, SPE Reservoir Evaluation \& Engineering, 4 (2001), pp.~146--160.

\bibitem{OPGEE_v3}
{\sc M.~Brandt, A.R.and~Masnadi, J.~Rutherford, H.~El-Houjeiri, Q.~Langfitt, K.~Vafi, C.~Yulia, and J.~Duffy}, {\em {OPGEE v3.0b User Guide And Technical Documentation}},  (2022).

\bibitem{Caterpillar2012}
{\sc Caterpillar}, {\em Oil \& gas production power spec sheets}, tech. rep., Caterpillar Oil \& Gas Resource Center, 2012.

\bibitem{Caulton2014a}
{\sc D.~R. Caulton, P.~B. Shepson, M.~O. Cambaliza, D.~McCabe, E.~Baum, and B.~H. Stirm}, {\em Methane destruction efficiency of natural gas flares associated with shale formation wells}, Environmental science \& technology, 48 (2014), pp.~9548--9554.

\bibitem{Chambers2003}
{\sc Chambers}, {\em Well test flare plume monitoring phase ii: Dial testing in alberta}, Tech. Rep. Project No: CEM 7454-2003, Alberta Research Council, December 2003.

\bibitem{Dillon2003}
{\sc J.~Dillon}, {\em Treatment technology review and assessment for petroleum process waters}, PhD thesis, Imperial College, 2003.

\bibitem{Dixit2023}
{\sc Y.~Dixit et~al.}, {\em Carbon intensity of global crude oil trading and market policy implications}, Nature Communications,  (2023).

\bibitem{Hassan2013}
{\sc H.~M. El-Houjeiri, A.~R. Brandt, and J.~E. Duffy}, {\em Open-source lca tool for estimating greenhouse gas emissions from crude oil production using field characteristics}, Environmental Science \& Technology, 47 (2013), pp.~5998--6006.
\newblock PMID: 23634761.

\bibitem{Elvidge2009}
{\sc C.~Elvidge, D.~Ziskin, K.~Baugh, B.~Tuttle, T.~Ghosh, D.~Pack, E.~Erwin, and M.~Zhizhin}, {\em A fifteen year record of global natural gas flaring derived from satellite data}, Energies, 2 (2009), pp.~595--622.

\bibitem{Fanchi2007}
{\sc J.~R. Fanchi}, {\em Volume i: General engineering}, in Petroleum Engineering Handbook, L.~W. Lake, ed., Society of Petroleum Engineers, Richardson, TX, 2007.

\bibitem{GE2011}
{\sc GE}, {\em Ge motors catalog 1.3}, tech. rep., General Electric, 2011.

\bibitem{Gvakharia2017}
{\sc A.~Gvakharia, E.~A. Kort, A.~Brandt, J.~Peischl, T.~B. Ryerson, J.~P. Schwarz, M.~L. Smith, and C.~Sweeney}, {\em Methane, black carbon, and ethane emissions from natural gas flares in the bakken shale, north dakota}, Environmental Science \& Technology, 51 (2017), pp.~5317--5325.

\bibitem{Hendler2009}
{\sc A.~Hendler, J.~Nunn, and J.~Lundeen}, {\em Voc emissions from oil and condensate storage tanks}, tech. rep., Texas Environmental Research Consortium, 2009.

\bibitem{Johnson2008}
{\sc M.~Johnson}, {\em Flare efficiency \& emissions: Past \& current research}, in Global Forum on Flaring and Venting Reduction and Natural Gas Utilisation, 2008.

\bibitem{Johnson2001}
{\sc M.~Johnson, D.~Wilson, and L.~Kostiuk}, {\em A fuel stripping mechanism for wake-stabilized jet diffusion flames in crossflow}, Combustion Science and Technology, 169 (2001), pp.~155--174.

\bibitem{koller2009probabilistic}
{\sc D.~Koller and N.~Friedman}, {\em {Probabilistic Graphical Models: Principles and Techniques}}, MIT Press, 2009.

\bibitem{Liang2020}
{\sc J.~Liang et~al.}, {\em Carbon intensity of global crude oil refining and mitigation potential}, Nature Climate,  (2020).

\bibitem{masnadi2018}
{\sc M.~S. Masnadi et~al.}, {\em Global carbon intensity of crude oil production}, Science, 361 (2018), pp.~851--853.

\bibitem{Masnadi2021}
\leavevmode\vrule height 2pt depth -1.6pt width 23pt, {\em Carbon implications of marginal oils from market-derived demand shocks}, Science,  (2021).

\bibitem{MASNADI_proxy}
{\sc M.~S. Masnadi, P.~R. Perrier, J.~Wang, J.~Rutherford, and A.~R. Brandt}, {\em Statistical proxy modeling for life cycle assessment and energetic analysis}, Energy, 194 (2020), p.~116882.

\bibitem{Ozumba2000}
{\sc C.~Ozumba and I.~Okoro}, {\em Combustion efficiency measurements of flares operated by an operating company}, SPE International Conference on Health, Safety, and the Environment in Oil and Gas Exploration and Production,  (2000).

\bibitem{Pearl1988}
{\sc J.~Pearl}, {\em Probabilistic reasoning in intelligent systems: Networks of beliefs}, Morgan Kaufmann Publishers, Inc., 1988.

\bibitem{Rutherford2021}
{\sc J.~S. Rutherford, E.~D. Sherwin, A.~P. Ravikumar, G.~A. Heath, J.~Englander, D.~Cooley, D.~Lyon, M.~Omara, Q.~Langfitt, and A.~R. Brandt}, {\em Closing the methane gap in us oil and natural gas production emissions inventories}, Nature Communications, 12 (2021).

\bibitem{IPCC2007}
{\sc S.~Solomon, D.~Qin, M.~Manning, Z.~Chen, M.~Marquis, K.~B. Averyt, M.~Tignor, and H.~L. Miller}, eds., {\em Contribution of Working Group I to the Fourth Assessment Report of the Intergovernmental Panel on Climate Change 2007}, Cambridge University Press, 2007.

\bibitem{IPCC2013}
{\sc T.~Stocker, D.~Qin, G.-K. Plattner, M.~Tignor, S.~Allen, J.~Boschung, A.~Nauels, Y.~Xia, V.~Bex, and P.~Midgley}, eds., {\em Climate Change 2013: The Physical Science Basis.}, Cambridge University Press, Cambridge, United Kingdom, 2013.

\bibitem{Solarturbines2012}
{\sc S.~Turbines}, {\em Gas turbine generator sets}, tech. rep., Solar Turbines, 2012.

\bibitem{2020SciPy-NMeth}
{\sc P.~Virtanen, R.~Gommers, T.~E. Oliphant, M.~Haberland, T.~Reddy, D.~Cournapeau, E.~Burovski, P.~Peterson, W.~Weckesser, J.~Bright, S.~J. {van der Walt}, M.~Brett, J.~Wilson, K.~J. Millman, N.~Mayorov, A.~R.~J. Nelson, E.~Jones, R.~Kern, E.~Larson, C.~J. Carey, {\.I}.~Polat, Y.~Feng, E.~W. Moore, J.~{VanderPlas}, D.~Laxalde, J.~Perktold, R.~Cimrman, I.~Henriksen, E.~A. Quintero, C.~R. Harris, A.~M. Archibald, A.~H. Ribeiro, F.~Pedregosa, P.~{van Mulbregt}, and {SciPy 1.0 Contributors}}, {\em {{SciPy} 1.0: Fundamental Algorithms for Scientific Computing in Python}}, Nature Methods, 17 (2020), pp.~261--272.

\bibitem{Vlasopoulos2006}
{\sc N.~Vlasopoulos, F.~Memon, D.~Butler, and R.~Murphy}, {\em Life cycle assessment of wastewater treatment technologies treating petroleum process waters}, Science of the Total Environment, 367 (2006), pp.~58--70.

\bibitem{Wang2009}
{\sc M.~Wang}, {\em Ca-greet model v1.8b}, computer program, Argonne National Laboratory, 2009.

\bibitem{Zavala2021}
{\sc D.~Zavala-Araiza, M.~Omara, R.~Gautam, M.~L. Smith, S.~Pandey, I.~Aben, V.~Almanza-Veloz, S.~Conley, S.~Houweling, E.~A. Kort, et~al.}, {\em A tale of two regions: methane emissions from oil and gas production in offshore/onshore mexico}, Environmental Research Letters, 16 (2021), p.~024019.

\end{thebibliography}
\bibliographystyle{siam}

\end{document}